\documentclass[12pt]{article}
\usepackage{geometry} 
\usepackage{amssymb}
\usepackage{amsmath}
\usepackage{MnSymbol}
\usepackage{hyperref}
\usepackage{braket}
\usepackage{cite}
\usepackage{subcaption,tikz}
\usetikzlibrary{arrows,decorations.markings,decorations.pathreplacing}

\newcommand{\be}{\begin{equation}}
\newcommand{\ee}{\end{equation}}
\newcommand{\bea}{\begin{eqnarray}}
\newcommand{\eea}{\end{eqnarray}}

\renewcommand{\k}{\kappa}

\newcommand{\C}{\mathbb{C}}

\newcommand{\Z}{\mathbb{Z}}

\newcommand{\mcO}{\mathcal{O}}

\tikzset{
	partial ellipse/.style args={#1:#2:#3}{
		insert path={+ (#1:#3) arc (#1:#2:#3)}
	}
}

\numberwithin{equation}{section}

\begin{document}
	
\begin{flushright}
	MI-HET-787
\end{flushright}

	
\begin{center}

{\large\bf Decomposition, Trivially-Acting Symmetries, and Topological Operators}

	\vspace*{0.2in}
	
	Daniel G. Robbins$^1$, Eric Sharpe$^2$,
	Thomas Vandermeulen$^3$
	
	\vspace*{0.1in}
	
	\begin{tabular}{cc}
		{\begin{tabular}{l}
				$^1$ Department of Physics\\
				University at Albany\\
				Albany, NY 12222 \end{tabular}} &
		{\begin{tabular}{l}
				$^2$ Department of Physics MC 0435\\
				850 West Campus Drive\\
				Virginia Tech\\
				Blacksburg, VA  24061 \end{tabular}}
	\end{tabular}
		{\begin{tabular}{l}
				$^3$ George P.~and Cynthia W.~Mitchell Institute\\
				for Fundamental Physics and Astronomy\\
				Texas A\&M University\\
				College Station, TX 77843 \end{tabular}}
			
	\vspace*{0.2in}
	
	{\tt dgrobbins@albany.edu},
	{\tt ersharpe@vt.edu},
	{\tt tvand@tamu.edu}
	
\end{center}
\pagenumbering{gobble}

Trivially-acting symmetries in two-dimensional conformal field theory include twist fields of dimension zero which are local topological operators.  We investigate the consequences of regarding these operators as part of the global symmetry of the theory. That is, we regard such a symmetry as a mix of topological defect lines (TDLs) and topological point operators (TPOs).  TDLs related by a trivially-acting symmetry can join at a TPO to form non-trivial two-way junctions.  Upon gauging, the local operators at those junctions can become vacua in a disjoint union of theories.  Examining the behavior of the TPOs under gauging therefore allows us to refine decomposition by tracking the trivially-acting symmetries of each universe.  Mixed anomalies between the TDLs and TPOs provide discrete torsion-like phases for the partition functions of these orbifolds, modifying the resulting decomposition.  This framework also readily allows for the consideration of trivially-acting non-invertible symmetries.

\begin{flushleft}
November 2022
\end{flushleft}

\newpage

\tableofcontents

\newpage

\section{Introduction}
\label{sec:intro}
\pagenumbering{arabic}

A common assumption made in the study of global symmetries is that they act effectively on the states of a field theory.  The purpose of this work is to systematically explore the consequences of relaxing this assumption in a modern framework along the lines of \cite{Chang:2018iay,Gaiotto:2014kfa,Frohlich:2009gb,Brunner:2013ota} where symmetries are associated to extended objects, specifically nonlocal topological operators.  Fortunately, there already exist a number of results in this direction to build upon.  It has been long known that gauging trivially-acting symmetries in conformal field theory (CFT) leads to a violation of cluster decomposition, but in a mild manner: the resulting theories are equivalent to direct sums of local theories, a phenomenon known as decomposition.\footnote{For more details on how such theories violate cluster decomposition, see \cite[section 2]{Hellerman:2006zs}.  The fact that violating cluster decomposition in this manner is itself called decomposition may be confusing, but the term is by now standard in the literature.}

More formally, decomposition is the observation that in $d>1$ dimensions a local quantum field
theory with a $(d-1)$-form symmetry is equivalent to (decomposes into) a disjoint union of other theories, known as universes.  Decomposition was first described in 2006 
in \cite{Hellerman:2006zs} as part of efforts to understand string compactifications on generalizations of spaces known as stacks, where it resolved some of the apparent physical inconsistencies of those theories.  Since that time, it has been applied in a wide variety of areas, including Gromov-Witten theory (see e.g.~\cite{ajt1,ajt2,ajt3,t1,gt1,xt1}), to give nonperturbative constructions of geometries in gauged linear sigma models (see e.g.~\cite{Caldararu:2010ljp,Hori:2011pd,Sharpe:2012ji,Addington:2012zv,Halverson:2013eua,Hori:2013gga,Hori:2016txh,Wong:2017cqs,Kapustka:2017jyt,Chen:2018qww,Chen:2020iyo,Guo:2021aqj}), to understand IR limits of supersymmetric pure gauge theories (see e.g.~\cite{Eager:2020rra}), in two-dimensional adjoint QCD (see e.g.~\cite{Komargodski:2020mxz}), and in \cite{Robbins:2020msp,Robbins:2021lry,Robbins:2021ibx,Robbins:2021xce} to understand the  Wang-Wen-Witten anomaly resolution of \cite{Wang:2017loc}. See also e.g.~\cite{Hellerman:2010fv,Anderson:2013sia,Sharpe:2014tca,Sharpe:2019ddn,Tanizaki:2019rbk,Nguyen:2021yld,Nguyen:2021naa,Yu:2020twi,Cherman:2020cvw,Cherman:2021nox,Honda:2021ovk,Gu:2021yek,Gu:2021beo,Sharpe:2021srf,Meynet:2022bsg,Pantev:2022kpl,Pantev:2022pbf,Lin:2022xod} for other applications and tests in theories in dimensions two, three, and four, and see \cite{Sharpe:2006vd,Sharpe:2010zz,Sharpe:2010iv,Sharpe:2019yag,Sharpe:2022ene} for some reviews of the subject.

Prototypical examples of decomposition include gauge
theories and orbifolds in which a noneffectively-acting higher form
symmetry group is gauged, see e.g.\cite{Pantev:2005rh,Pantev:2005zs,Pantev:2005wj,Pantev:2022kpl,Pantev:2022pbf}.
However, these studies have so far largely focused on the decomposition of local operators.  In this paper, we will begin a study of how extended objects behave under decomposition.  For concreteness we will focus on two-dimensional theories, as that is the context in which decomposition has already been studied most heavily, so we can most easily relate the story told here to existing results.  That said, we expect the picture proposed here to generalize straightforwardly to the case of gauging a trivially-acting $(d-2)$-form symmetry in $d$ spacetime dimensions.

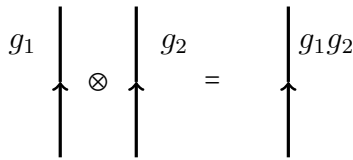
\begin{figure}
	\centering
	\begin{tikzpicture}
	\draw[very thick,->] (0,0) -- (0,1);
	\draw[very thick] (0,1) -- (0,2);
	\node at (-0.5,1.5) {$g_1$};
	\node at (0.5,1) {$\otimes$};
	\draw[very thick,->] (1,0) -- (1,1);
	\draw[very thick] (1,1) -- (1,2);
	\node at (1.5,1.5) {$g_2$};
	\node at (2,1) {$=$};
	\draw[very thick,->] (3,0) -- (3,1);
	\draw[very thick] (3,1) -- (3,2);
	\node at (3.5,1.5) {$g_1g_2$};
	\end{tikzpicture}
	\caption{The fusion of two group-like line operators.}
	\label{fig:linefuse}
\end{figure}

Throughout this paper we begin with a two-dimensional theory with an effectively-acting symmetry group $G$ (which we will assume finite and non-anomalous), generated by TDLs\footnote{For a physics-oriented review of symmetries as topological operators in two dimensions, see \cite{Chang:2018iay}.} which are labeled by elements of $G$ and fuse according to the group relations in $G$, as shown in Figure~\ref{fig:linefuse}.  We then extend the TDLs to a distinguished set describing a group extension $\Gamma$, 
\be
\label{gammaext}
1 \: \longrightarrow \: K \: \longrightarrow \:
 \Gamma \: \longrightarrow \: G \: \longrightarrow \: 1,
\ee
with $K$ acting trivially.  By trivially-acting we mean that an element $k\in K$ acts on any local operator $\mathcal{O}$ as
\be
\label{kaction}
k \cdot \mathcal{O} = \mathcal{O}
\ee
for all $k\in K$ and all local operators $\mathcal{O}$.  As we are in two dimensions, (\ref{gammaext}) and (\ref{kaction}) fully characterize the $K$ lines -- they act trivially on local operators and fuse with other topological line operators according to the $\Gamma$ group law.  Because trivially-acting symmetries are in this sense `invisible' to the spectrum of the theory, it is common to restrict one's attention exclusively to effective symmetries.  In this work, however, we will attempt to elucidate some of the structure carried by these trivially-acting symmetries and their interplay with a theory's effective symmetries.\footnote{We will also use the term `non-effectively-acting' symmetry to mean a symmetry such as $\Gamma$ for which only a subgroup acts trivially.}

Note that the group extension (\ref{gammaext}) describes the total zero-form symmetry of the system.  In section~\ref{sec:main} we will add local topological operators to the story, which can be associated with one-form symmetries.  One might then be tempted to assign a higher-group-type algebraic structure to the resulting operators -- appendix~\ref{app:mixedext} discusses this interpretation.  As many readers would be familiar with the concept of a 2-group (the extension of a zero-form symmetry by a one-form symmetry) it may be worth mentioning that the operators we consider here do not form such a structure.

Each TDL in the resulting theory\footnote{The underlying philosophy here is that a ``theory'' is defined not only by local data, but also by its non-local spectrum, as in \cite{Gaiotto:2014kfa}.} is labeled by an element of $\Gamma$, whose underlying set is that of $K\times G$.  When this extension is trivial, $\Gamma$ is in fact equal to $K\times G$, and each symmetry element is some (noninteracting) mix of the effective and trivial symmetries.  Possible non-trivial extensions 
describe how these symmetries can mix together, in that we could fuse two $G$ lines and end up with a $K$ line.  Note that non-trivial extensions will in general prevent $G$ from being a subgroup of $\Gamma$, which means that the theory under consideration may no longer have a standalone $G$ symmetry.

In ungauged theories, if a TDL describes a noneffectively-acting group, then the lines for the trivially-acting subgroup can end on TPOs (which are bound to the lines and do not define stand-alone one-form symmetries).  After gauging, those TPOs become unbound point operators generating one-form symmetries.  Tracking such point operators determines TDL structures, which enables one to follow TDLs and TPOs through gauging. In this fashion, we refine decomposition by explicitly tracking such extended objects.

After a review of the requisite basic notions in section~\ref{sec:background}, section~\ref{sec:main} introduces the notion that trivially-acting symmetries are controlled by a mix of topological line and point operators.  This perspective allows us to consider the effect of a mixed anomaly when gauging the line operators.  Alternatively, in the absence of such anomalies, we could simultaneously gauge the lines and points.  Equipped with this technology, section~\ref{sec:examples} runs through a handful of examples, the final of which incorporates non-invertible symmetries in the form of a trivially-acting categorical symmetry.  We will see that this framework allows us to track the fate of trivially-acting symmetries under decomposition, with the upshot being that they divide themselves among the various universes.

\section{Background}
\label{sec:background}

This section reviews the various concepts that will enter into our full treatment of trivially-acting symmetries.  The material presented here is not new, so readers who are familiar with these subjects may wish to skim the section for notation or simply skip ahead to section~\ref{sec:main}.

\subsection{Higher-Form Symmetries}

We have presented a notion of symmetries as actions on local quantities generated by topological operators of codimension one.  In field theory, however, we often consider extended objects in addition to local ones.  It would be natural to consider an analog of symmetries for these objects -- transformations between extended objects.  In fact, analogously to ordinary symmetries, transformations of $n$-dimensional objects are naturally described by operators of codimension $n+1$.  Such transformations are termed $n$-form symmetries.

One realization of higher-form symmetries involves `groups' in which properties such as associativity are weakened.  Higher-form symmetries have a long history in physics, e.g.~\cite{Freed:1994ad,Baez:2010ya,Sati:2010ss,Sati:2010dc,Sati:2008eg}, in discussions of the String group (see e.g.~\cite{Schommer-Pries:2011vyj,Baez:2005sn,Nikolaus:2011zg}), in the Yetter model (see e.g.~\cite{yet,martins-porter}), and in lattice gauge theories (see e.g.~\cite{Grosse:2001kw,Pfeiffer:2003je}), to name a few examples.  Additional categorical generalizations of (orbifold) groups, via an application of defects to generalize the ordinary orbifold construction, are discussed in \cite{Frohlich:2009gb,Carqueville:2012dk,Carqueville:2013mpa,Brunner:2013ota,Brunner:2014lua,Carqueville:2015pra}. See also \cite{Wang:2014pma} for related ideas and applications of discrete gauge theories and group cohomology in condensed matter physics.

In this paper, we will focus on two dimensions.  There, ordinary symmetries are controlled by TDLs.  There exists one type of higher-form symmetry: one-form symmetries, controlled by topological point operators (TPOs).  These describe how line operators transform into each other.  Already in two dimensions we have a fairly rich structure -- the generators of the two types of symmetries naturally act on each other.  This interplay, as we will see, will be key to fully describing trivially-acting symmetries.

\subsection{Non-Invertible Symmetries}

Broadly speaking, topological defect lines do not obey group laws, but rather more general fusion relations.  For instance, there is a non-invertible operator that implements Kramers-Wannier duality in the critical Ising model \cite{Frohlich:2004ef}.  Additionally, and more relevant to this paper, in order to fully capture the quantum dual to a non-abelian symmetry, one is forced to examine non-group-like symmetries \cite{Bhardwaj:2017xup}.

The fusion of general TDLs is described mathematically by unitary fusion categories; see \cite{Fuchs:2002cm,Bhardwaj:2017xup}.  In such a category, we assume that there exists a set of `simple' objects, and all other objects in the category are expressible as non-negative integral linear combinations of these simple objects.  There are a set of fusion rules which govern how the simple objects (and hence the composite ones) combine, which along with addition gives the structure of a ring.  There is a map $X\otimes(Y\otimes Z)\to (X\otimes Y)\otimes Z$ known as the associator, and a number of consistency conditions which will not be important for our purposes.

\begin{figure}
\centering
\begin{tabular}{| c | c | c | c |}
	\hline
	$\otimes$ & $1$ & $X$ & $Y$ \\\hline
	$1$ & $1$ & $X$ & $Y$ \\\hline
	$X$ & $X$ & $1$ & $Y$ \\\hline
	$Y$ & $Y$ & $Y$ & $1+X+Y$\\\hline
\end{tabular}
\caption{The fusion products for Rep$(S_3)$.}
\label{fig:reps3}
\end{figure}

A group $G$ defines a special case of this structure:  there is a simple line for each element $g\in G$, and the fusion rules are simply the group relations.  To describe anomalies, one can add a non-trivial associator, given by a representative of an element of $H^3(G,U(1))$.  Less trivial examples include quantum symmetries of orbifolds.  When we gauge a symmetry given by a group $G$, the gauged theory has a quantum $\text{Rep}(G)$ symmetry where simple lines can be labeled by irreducible representations of $G$.  Since the irreducible representations of a non-abelian group can have dimension greater than one, irreducible representations in general do not form groups under tensor products, but there does exist a fusion law for products of irreducible representations. 

An example which we will use repeatedly is the symmetric group $S_3$.  It has three irreducible representations: the trivial representation, a non-trivial representation of dimension one, and a single irreducible representation of dimension two.  Labeling these irreducible representations respectively as $1$, $X$ and $Y$, their fusion products are as shown in Figure~\ref{fig:reps3}.  In particular, the simple line $Y$ corresponding to the two-dimensional irrep has the fusion relation $Y\otimes Y=1+X+Y$ -- in the group-like case, fusion of simple lines never produces a sum.  Correspondingly, one can see that there is no object $Y^{-1}$ satisfying $Y\otimes Y^{-1}=1$, hence the term `non-invertible.'

\subsection{Gauging Symmetries}

We will use
a notion of gauging which is sufficiently general to include the categorical symmetries described above.  
In order to gauge a general symmetry, we select an object $A$,
known as the algebra object,
along with suitable multiplication and identity morphisms to define a symmetric Frobenius algebra \cite{Frohlich:2004ef,Bhardwaj:2017xup}.\footnote{In the condensed matter literature, this generalized form of gauging is known as anyon condensation \cite{Hung:2015hfa}.}  

Intuitively $A$ serves as the identity operator for the gauged theory.  We calculate quantities in the gauged theory from ungauged objects by inserting a ``sufficiently fine mesh'' of $A$ lines, along with appropriate normalization.  As an example, when gauging a group-like symmetry, we can take $A$ to be the sum of lines labeled by elements of the group, i.e.
\be
A \: = \: \sum_{g\in G}L_g
\ee
where $L_g$ denotes a TDL labeled by $g$.  In order to calculate e.g.~the torus partition function in the gauged theory, we wrap both cycles with an $A$ line and normalize by the dimension of $A$, which will simply be the order of the group.  Breaking the two $A$ lines into sums over group elements, this reduces to the familiar prescription for the orbifold torus partition function:
\be
\frac{1}{|A|}Z_{A,A}
\: = \:
\frac{1}{|G|}\sum_{\substack{g_1,g_2\in G \\ [g_1,g_2]=1}}Z_{g_1,g_2}.
\ee
The advantage to such a general formalism is that we can also gauge non-group-like symmetries.  For example, returning to Rep$(S_3)$, there exist suitable choices of multiplication and unit morphisms such that $1+X$, $1+Y$ and $1+X+2Y$ constitute algebra objects.  Taking $A=1+X$ gives us a group-like $\Z_2$ gauging, while $1+Y$ and $1+X+2Y$ lead to non-group-like gaugings (at least superficially -- these two gaugings are `dual' to gauging $\Z_3\subset S_3$ and all of $S_3$, respectively).

The algebra object also has a role to play in determining the local operators that appear in the gauged theory -- in particular we will be interested in operators of weight zero.  Acting with $A$ on TPOs will have the effect of ``projecting out'' linear combinations that do not survive as bulk local operators, such that the result is a linear combination of TPOs which are freestanding in the gauged theory.

\subsection{Decomposition}
\label{sec:decomp}

In this section, we will review decomposition in two-dimensional orbifold CFTs.
Decomposition for two-dimensional orbifolds without discrete torsion can be summarized as follows \cite{Hellerman:2006zs,Robbins:2020msp}.  Let $T$ be a theory with an action of $\Gamma$, where $K \subset \Gamma$ acts trivially and $G = \Gamma/K$.  Assume that there is no discrete torsion in the $\Gamma$ orbifold.
(Decomposition for two-dimensional orbifolds including discrete torsion is described in \cite{Robbins:2020msp}, with mixed anomalies in \cite{Robbins:2021ibx}, and other miscellaneous two-dimensional orbifold cases
are in \cite{Robbins:2021lry}.)  Then, \begin{equation}  \label{eq:decomp:genl}{\rm QFT}\left( [T/\Gamma] \right) \: = \:
{\rm QFT} \left( \left[ \frac{ T \times \hat{K} }{G} \right]_{\hat{\omega}} \right),\end{equation} where in general the effectively-acting coset $G$ acts on both $T$ and $\hat{K}$, and $\hat{\omega}$ denotes discrete torsion appearing in the universes, as explained in \cite{Hellerman:2006zs,Robbins:2020msp}.  The right-hand-side is a disjoint union of theories, as many as orbits of $G$ on $\hat{K}$.  The elements of that disjoint union are known as the universes of the decomposition.

For  later use, it will be helpful to give explicitly the projectors\footnote{The state-operator correspondence maps the vacuum state in each universe to an operator which projects onto that universe.  Due to this mapping, we will be somewhat abusive with notation and use $\Pi$ to refer to both the vacuum state and the projection operator.} onto
the universes of decomposition, referring specifically to decomposition in two-dimensional orbifolds without discrete torsion, as described in~(\ref{eq:decomp:genl}).  These are essentially a consequence
of Wedderburn's theorem, and are given explicitly in \cite[section 2.2.2]{Sharpe:2021srf} in terms of TPOs.  
For $\Gamma$ orbifolds without discrete torsion, we give them as follows.  Let $R$ be a representation of $K$ associated to a universe, meaning that 
\begin{equation}
R \: = \: \bigoplus_i R_i
\end{equation}
for $R_i \in \hat{K}$ the irreducible representations forming an orbit of $G$.  Then,
\begin{equation} 
\label{eq:proj}
\Pi_R \: = \: \sum_i \Pi_{R_i},
\end{equation}
where
\begin{equation}
\label{projformula}
\Pi_{R_i} \: = \: \frac{ \dim R_i }{ |K| }
\sum_{k \in K} \chi_{R_i}(k^{-1}) \, \sigma_k,
\end{equation}
and where $\chi_{R_i}$ denotes a character of $R_i$ and the $\sigma_k$ are ungauged local operators which fuse according to the group law in $K$.\\

To make this paper self-contained, it may be helpful to briefly illustrate decomposition in some simple examples.
Let $T$ denote a theory, and $K$ an abelian group with a trivial action on $T$.  In terms of the prediction~(\ref{eq:decomp:genl}), $G=1$ so $\Gamma = K$ and decomposition predicts in this case that
\begin{equation}
{\rm QFT}\left( [T/K] \right) \: = \: {\rm QFT}\left( T \times \hat{K} \right),
\end{equation}
as many copies of $T$ as irreducible representations of $K$, namely $|K|$ since $K$ is abelian.  We can check this by computing partition functions.  In calculating the partition function for the orbifold $[T/K]$, each sector $Z_{k_1,k_2}$ is simply going to be equivalent to the parent theory partition function.  Therefore we have
\be
Z_{[T/K]} \: = \:
\frac{1}{|K|}\sum_{k_1,k_2\in K}Z_{k_1,k_2}
\: = \: \frac{1}{|K|}\sum_{k_1,k_2\in K}Z_T
\: = \: |K|Z_T,
\ee
which confirms the statement of decomposition in this case.

As with orbifolds by effective symmetries, there is a quantum symmetry in this orbifold which we can gauge to return the original theory, given by $\hat{K}$.  In this context we can interpret $\hat{K}$ as a subgroup\footnote{By virtue of Cayley's theorem.} of the exchange symmetry $S_{|K|}$ acting on the $|K|$ copies appearing
in the decomposition.

If $K$ is non-abelian, a similar story applies, and the orbifold $[T/K]$ is equivalent to as many copies of $T$ as irreducible representations of $K$.  As before, we can check using partition functions.  Partition functions on $T^2$ are computed by summing over commuting pairs, which gives
\be
\label{nonabdecomp}
Z_{[T/K]} \: = \:
 \frac{1}{|K|}\sum_{\substack{k_1,k_2\in K \\ [k_1,k_2]=1}}Z_{k_1,k_2}
\: = \: N_KZ_T,
\ee
where $N_K$ is the number of conjugacy classes in $K$.  There is once again a quantum symmetry we can gauge to return to the original theory, except it is now given by the fusion category $\text{Rep}(K)$ described by the fusion structure of the irreducible representations of $K$, which will fail to be group-like due to the presence of irreducible representations with dimension greater than one.\\

As an example, consider orbifolding a theory by the symmetric group $S_3$, acting trivially (so that $G=1$ and $\Gamma = K=S_3$).  This group has three conjugacy classes, hence three irreducible representations, labeled $1$, $X$, $Y$ earlier, so from (\ref{nonabdecomp}) and (\ref{eq:decomp:genl}) (and the fact that the effectively-acting coset is $G=1$) we expect three universes (i.e.~a decomposition into a direct sum of three theories).  

We expect there to be a quantum $\text{Rep}(S_3)$ symmetry acting on the vacua, the fusion products for which were presented in Figure~\ref{fig:reps3}.  We can see this as follows.  First, label the projectors into these universes as $\Pi_1$, $\Pi_X$, $\Pi_Y$.  The quantum symmetry defines a product of the form
\begin{equation}
R \cdot \sigma_k \: = \: \chi_R(k) \, \sigma_k,
\end{equation}
for a representation $R$ and $k \in S_3$ (where $\sigma_k$ corresponds to an ungauged local operator), which can be applied to compute
\begin{equation}
1 \Pi_a \: = \: \Pi_a \: \: \mbox{ for all }a,
\end{equation}
\begin{equation}  \label{reps3-1}
X \Pi_1 \: = \: \Pi_X, \: \: \: X \Pi_X \: = \: \Pi_1,
\end{equation}
\begin{equation} \label{reps3-2}
X \Pi_Y \: = \: \Pi_Y,
\end{equation} 
\begin{equation}  \label{reps3-3}
Y \Pi_1 \: = \: \frac{1}{2} \Pi_Y \: = \: Y \Pi_X,
\end{equation}
\begin{equation} \label{reps3-4}
Y \Pi_Y \: = \: 2 \Pi_1 \: + \: 2 \Pi_X \: + \: \Pi_Y.
\end{equation}
We will give explicit expressions for the projectors and other details later in section~\ref{sect:ex:s3}.

Here we see a qualitative difference between invertible and non-invertible symmetries.  The $\Z_2$ subsymmetry generated by $X$ exchanges two of the three copies, acting on the sum of the universes as $X(\Pi_1+\Pi_X+\Pi_Y)=(\Pi_1+\Pi_X+\Pi_Y)$.  The weight two line $Y$, however, has nonintegral action on the individual universes, and acts on their sum as $Y(\Pi_1+\Pi_X+\Pi_Y)=2(\Pi_1+\Pi_X+\Pi_Y)$.

We have illustrated decomposition in a handful of simple examples to make this paper self-contained.  Many more examples in two-dimensional orbifolds are discussed in \cite{Hellerman:2006zs,Robbins:2020msp,Robbins:2021lry,Robbins:2021ibx,Robbins:2021xce,Lin:2022xod}.

\section{Trivially-Acting Symmetries}
\label{sec:main}

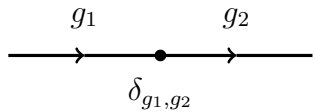
\begin{figure}[h]
	\centering
	\begin{tikzpicture}
	\draw[very thick,->] (0,0) -- (1,0);
	\draw[very thick] (1,0) -- (2,0);
	\draw[very thick,->] (2,0) -- (3,0);
	\draw[very thick] (3,0) -- (4,0);
	\node at (1,0.5) {$g_1$};
	\node at (3,0.5) {$g_2$};
	\filldraw[black] (2,0) circle (2pt);
	\node at (2,-0.5) {$\delta_{g_1,g_2}$};
	\end{tikzpicture}
	\caption{Two effective symmetry lines joined at a topological junction.}
	\label{fig:linepoint_eff}
\end{figure}

We begin the analysis of extended operators and trivially-acting symmetries by looking at junctions between two TDLs, focusing on the topological point operators which live there.  In the effective, group-like case, this relatively uninteresting configuration is pictured in Figure~\ref{fig:linepoint_eff}.  Because junctions between lines are constrained by the group action, the only weight zero operator that can sit at a junction between a $g_1$ and $g_2$ line is the identity, and only when $g_1=g_2$.  Otherwise such a junction can only exist if it contains a higher weight operator.\footnote{To expand on this point, one often considers twist fields on which TDLs can end.  In the effective case, these are local operators, but are not topological.  Our main concern in this work is to point out that when the associated symmetry acts trivially, there will be twist fields of weight zero, which are therefore local topological operators, which one can regard as being part of the global symmetry of the theory.}

\begin{figure}[h]
	\centering
	\begin{tikzpicture}
	\draw[very thick,->] (0,0) -- (1,0);
	\draw[very thick] (1,0) -- (2,0);
	\draw[very thick,->] (2,0) -- (3,0);
	\draw[very thick] (3,0) -- (4,0);
	\node at (1,0.5) {$\gamma$}; 
	\node at (3,0.5) {$k\gamma$};  
	\filldraw[black] (2,0) circle (2pt);
	\node at (2,-0.5) {$\sigma_k$};
	\end{tikzpicture}
	\caption{Two line operators joined by a point operator.}
	\label{fig:linepoint}
\end{figure}
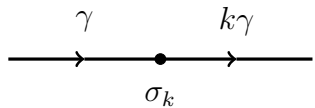

In the presence of trivially-acting symmetries, such junctions can support non-trivial weight zero operators.  More precisely, let the total zero-form symmetry of our theory be $\Gamma=K.G$\footnote{Here we are using the notation of \cite{ATLAS} where a dot indicates a general group extension extension.  That is, $K$ is a normal subgroup of $\Gamma$ and $\Gamma/K\simeq G$, but $K$ need not be central and the extension need not be split.}, where the normal subgroup $K$ acts trivially.  Elements of $\Gamma$ that differ by an element of $K$ can be joined at topological junctions, as shown in Figure~\ref{fig:linepoint}.  Similarly, a line labeled by an element of the trivially-acting subgroup $K$ can end at a corresponding point operator (which is to say it can have a two-way junction with the identity TDL).  While the information is redundant with the labelings of the lines, we will find it convenient to label such point operators by elements of $K$.

\begin{figure}[h]
	\centering
	\begin{tikzpicture}
	\draw[very thick,->] (0,0) -- (1,0);
	\draw[very thick] (1,0) -- (2,0);
	\draw[very thick,->] (2,0) -- (3,0);
	\draw[very thick] (3,0) -- (4,0);
	\node at (1,0.5) {$\gamma$}; 
	\node at (3,0.5) {$k_1\gamma$}; 
	\node at (5,0.5) {$k_2k_1\gamma$}; 
	\filldraw[black] (2,0) circle (2pt);
	\filldraw[black] (4,0) circle (2pt);
	\node at (2,-0.5) {$\sigma_{k_1}$};
	\node at (4,-0.5) {$\sigma_{k_2}$};
	\draw[very thick,->] (4,0) -- (5,0);
	\draw[very thick] (5,0) -- (6,0);
	\draw[thick,->] (3,-1) -- (3,-1.5);
	\draw[very thick,->] (0,-2.5) -- (1.5,-2.5);
	\draw[very thick] (1.5,-2.5) -- (3,-2.5);
	\draw[very thick,->] (3,-2.5) -- (4.5,-2.5);
	\draw[very thick] (4.5,-2.5) -- (6,-2.5);
	\node at (2,-2) {$\gamma$}; 
	\node at (4,-2) {$k_2k_1\gamma$}; 
	\filldraw[black] (3,-2.5) circle (2pt);
	\node at (3,-3) {$\sigma_{k_2 k_1}$};
	\end{tikzpicture}	
	\caption{The fusion of two point operators.}
	\label{fig:grouplaw}
\end{figure}
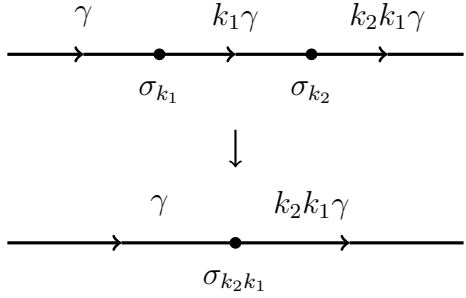

In the same way that the TDLs of our theory satisfy a fusion algebra (the group laws in the group-like case or a unitary fusion category in the more general case), the TPOs associated with these trivial symmetries can be given a fusion algebra structure.  Figure~\ref{fig:grouplaw} illustrates this -- the TPOs there labeled by $\sigma_{k_1}$ and $\sigma_{k_2}$ can be brought together, and we can use the operator product expansion (OPE) to fuse them into a single point operator joining the $\gamma$ and $k_2 k_1\gamma$ lines.  As we might have expected, the $\sigma_{k_1}$ and $\sigma_{k_2}$ operators fuse to the $\sigma_{k_2 k_1}$ operator, and in general the fusion rules for these TPOs are simply the group law in $K$.

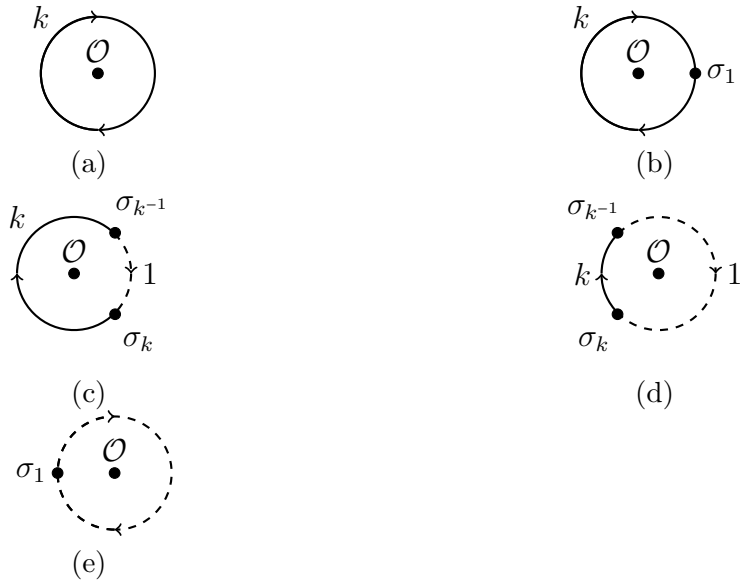
\begin{figure}
	\begin{subfigure}{0.5\textwidth}
		\centering
		\begin{tikzpicture}
		\draw[thick,->] (0,0) [partial ellipse=270:90:0.75cm and 0.75cm];
		\draw[thick,->] (0,0) [partial ellipse=270:-90:0.75cm and 0.75cm];
		\filldraw[black] (0,0) circle (2pt);
		\node at (0,0.3) {$\mathcal{O}$};
		\node at (-0.75,0.75) {$k$};
		\end{tikzpicture}
		\caption{}
		\label{fig:trivaction1}
	\end{subfigure}
	\begin{subfigure}{0.5\textwidth}
		\centering
		\begin{tikzpicture}
		\draw[thick,->] (0,0) [partial ellipse=270:90:0.75cm and 0.75cm];
		\draw[thick,->] (0,0) [partial ellipse=270:-90:0.75cm and 0.75cm];
		\filldraw[black] (0,0) circle (2pt);
		\node at (0,0.3) {$\mathcal{O}$};
		\node at (-0.75,0.75) {$k$};
		\filldraw[black] (0.75,0) circle (2pt);
		\node at (1.1,0) {$\sigma_1$};
		\end{tikzpicture}
		\caption{}
		\label{fig:trivaction2}
	\end{subfigure}
	\begin{subfigure}{0.5\textwidth}
		\centering
		\begin{tikzpicture}
		\draw[thick] (0,0) [partial ellipse=180:45:0.75cm and 0.75cm];
		\draw[thick,->] (0,0) [partial ellipse=-45:-180:0.75cm and 0.75cm];
		\draw[thick,dashed,->] (0,0) [partial ellipse=45:0:0.75cm and 0.75cm];
		\draw[thick,dashed] (0,0) [partial ellipse=0:-45:0.75cm and 0.75cm];
		\filldraw[black] (0,0) circle (2pt);
		\node at (0,0.3) {$\mathcal{O}$};
		\node at (-0.75,0.75) {$k$};
		\filldraw[black] (0.54,0.54) circle (2pt);
		\filldraw[black] (0.54,-0.54) circle (2pt);
		\node at (0.9,0.9) {$\sigma_{k^{-1}}$};
		\node at (1,0) {$1$};
		\node at (0.85,-0.9) {$\sigma_k$};
		\end{tikzpicture}
		\caption{}
		\label{fig:trivaction3}
	\end{subfigure}
	\begin{subfigure}{0.5\textwidth}
		\centering
		\begin{tikzpicture}
		\draw[thick] (0,0) [partial ellipse=180:135:0.75cm and 0.75cm];
		\draw[thick,->] (0,0) [partial ellipse=-135:-180:0.75cm and 0.75cm];
		\draw[thick,dashed,->] (0,0) [partial ellipse=135:0:0.75cm and 0.75cm];
		\draw[thick,dashed] (0,0) [partial ellipse=0:-135:0.75cm and 0.75cm];
		\filldraw[black] (0,0) circle (2pt);
		\node at (0,0.3) {$\mathcal{O}$};
		\node at (-1,0) {$k$};
		\filldraw[black] (-0.54,0.54) circle (2pt);
		\filldraw[black] (-0.54,-0.54) circle (2pt);
		\node at (-0.85,0.85) {$\sigma_{k^{-1}}$};
		\node at (1,0) {$1$};
		\node at (-0.85,-0.9) {$\sigma_k$};
		\end{tikzpicture}
		\caption{}
		\label{fig:trivaction4}
	\end{subfigure}
	\begin{subfigure}{0.5\textwidth}
		\centering
		\begin{tikzpicture}
		\draw[thick,dashed,->] (0,0) [partial ellipse=270:90:0.75cm and 0.75cm];
		\draw[thick,dashed,->] (0,0) [partial ellipse=270:-90:0.75cm and 0.75cm];
		\filldraw[black] (0,0) circle (2pt);
		\node at (0,0.3) {$\mathcal{O}$};
		\filldraw[black] (-0.75,0) circle (2pt);
		\node at (-1.1,0) {$\sigma_1$};
		\end{tikzpicture}
		\caption{}
		\label{fig:trivaction5}
	\end{subfigure}
	\caption{The action of $K$ on local operators is equivalent to a trivial action.}
	\label{fig:trivaction}
\end{figure}

We can see how such a configuration of point and line operators describes a trivially-acting symmetry by looking at the action on local operators, as in Figure~\ref{fig:trivaction}.  Figure~\ref{fig:trivaction1} depicts a line labeled by $k$ acting on a local operator.  In Figure~\ref{fig:trivaction2} we insert an identity operator, which in Figure~\ref{fig:trivaction3} we split into a $k$ and $k^{-1}$ pair\footnote{One might object that in moving from Figure~\ref{fig:trivaction1} to Figure~\ref{fig:trivaction5} there is the potential to pick up a $k$-dependent phase.  Such a phase should only arise from a gauge anomaly in $K$ \cite{Chang:2018iay}, and we are assuming here that there are no such anomalies in our trivially-acting symmetries.}, between which runs an identity line.  In Figure~\ref{fig:trivaction4} we bring these operators together from the other side, and finally in Figure~\ref{fig:trivaction5} we re-fuse them into an identity operator.  The remaining identity line and point operators can be removed, and the net result of this process is that $k\cdot\mathcal{O} = \mathcal{O}$, i.e.~$K$ acts trivially.

In passing, the computations of this section are formally similar to computations in \cite[section 4]{Bhardwaj:2022lsg}, although we believe our use differs.

\subsection{Mixed Anomalies}
\label{sec:mixedanom}

As we have discussed, line operators naturally act on local operators.  In this section we focus on the action of our TDLs (labeled by $\Gamma$) on our TPOs (labeled by $K$).  We will write these TPOs as $\sigma_k$.  There is a natural action of $\Gamma$ on these objects: conjugation.  However, there exists the possibility of adding non-trivial phases into this action.  We can make an ansatz that the general form of $\Gamma$ acting on $\sigma_k$ should be
\be
\label{bphase}
\gamma\cdot\sigma_k \: = \: \sigma_{\gamma k\gamma^{-1}}B(\gamma,\gamma k\gamma^{-1})
\ee
where $B(\gamma,k)\in U(1)$ is known as a mixed anomaly between the point and line operators.  

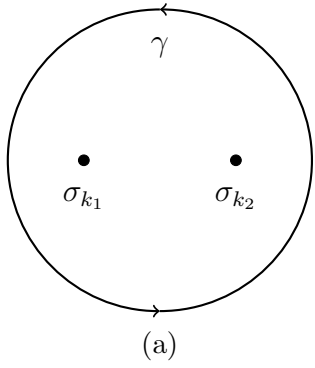
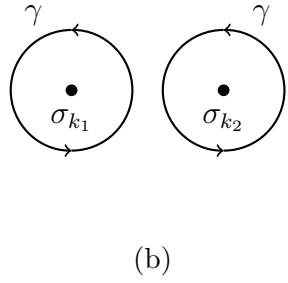
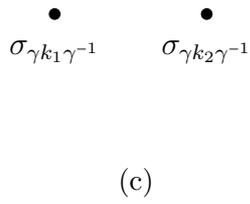
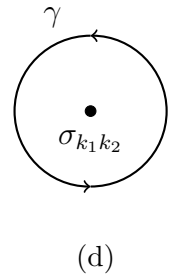
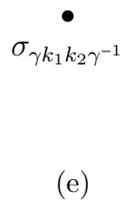
\begin{figure}
	\begin{subfigure}{0.5\textwidth}
		\centering
		\begin{tikzpicture}
		\filldraw[black] (-1,0) circle (2pt);
		\node at (-1,-0.5) {$\sigma_{k_1}$};
		\filldraw[black] (1,0) circle (2pt);
		\node at (1,-0.5) {$\sigma_{k_2}$};
		\draw[thick,->] (0,0) [partial ellipse=-90:90:2cm and 2cm];
		\draw[thick,->] (0,0) [partial ellipse=90:270:2cm and 2cm];
		\node at (0,1.5) {$\gamma$};
		\end{tikzpicture}
	\caption{}
	\label{homka}
	\end{subfigure}
	\begin{subfigure}{0.5\textwidth}
		\centering
		\begin{tikzpicture}
		\filldraw[black] (-1,0) circle (2pt);
		\node at (-1,-0.4) {$\sigma_{k_1}$};
		\filldraw[black] (1,0) circle (2pt);
		\node at (1,-0.4) {$\sigma_{k_2}$};
		\draw[thick,->] (-1,0) [partial ellipse=-90:90:0.8cm and 0.8cm];
		\draw[thick,->] (-1,0) [partial ellipse=90:270:0.8cm and 0.8cm];
		\node at (-1.5,1) {$\gamma$};
		\draw[thick,->] (1,0) [partial ellipse=-90:90:0.8cm and 0.8cm];
		\draw[thick,->] (1,0) [partial ellipse=90:270:0.8cm and 0.8cm];
		\node at (1.5,1) {$\gamma$};
		\end{tikzpicture}
		\vspace{1cm}
	\caption{}
	\label{homkb}
	\end{subfigure}
	\begin{subfigure}{0.5\textwidth}
		\centering
		\vspace{1cm}
		\begin{tikzpicture}
		\filldraw[black] (-1,0) circle (2pt);
		\node at (-1,-0.5) {$\sigma_{\gamma k_1\gamma^{-1}}$};
		\filldraw[black] (1,0) circle (2pt);
		\node at (1,-0.5) {$\sigma_{\gamma k_2\gamma^{-1}}$};
		\end{tikzpicture}
		\vspace{1cm}
	\caption{}
	\label{homkc}
	\end{subfigure}
	\begin{subfigure}{0.5\textwidth}
		\centering
		\vspace{0.5cm}
		\begin{tikzpicture}
		\filldraw[black] (0,0) circle (2pt);
		\node at (0,-0.4) {$\sigma_{k_1k_2}$};
		\draw[thick,->] (0,0) [partial ellipse=-90:90:1cm and 1cm];
		\draw[thick,->] (0,0) [partial ellipse=90:270:1cm and 1cm];
		\node at (-0.5,1.25) {$\gamma$};
		\end{tikzpicture}
		\vspace{0.5cm}
	\caption{}
	\label{homkd}
	\end{subfigure}
	\begin{subfigure}{0.5\textwidth}
		\centering
		\vspace{1cm}
		\begin{tikzpicture}
		\filldraw[black] (0,0) circle (2pt);
		\node at (0,-0.5) {$\sigma_{\gamma k_1k_2\gamma^{-1}}$};
		\end{tikzpicture}
		\vspace{1cm}
	\caption{}
	\label{homke}
	\end{subfigure}
	\caption{The fusion of two TPOs acted on by a TDL constrains the dependence of $B$ on its $K$ argument.}
	\label{fig:homk}
\end{figure}

Compatibility between fusion of TDLs and TPOs will constrain these phases.  More specifically, Figure~\ref{fig:homk} illustrates these consistency conditions for the $K$ argument.  In Figure~\ref{homka} we begin with two TPOs labeled by $k_1$ and $k_2$, circled by a TDL labeled by $\gamma$.  We can shrink the TDL and pinch it off into a circle surrounding each TPO, as in Figure~\ref{homkb}.  Each of these circles can be further shrunk to give the action of $\gamma$ on each point, leading to Figure~\ref{homkc}.  From (\ref{bphase}), Figures~\ref{homkb} and \ref{homkc} should be related to each other by a factor of $B(\gamma,\gamma k_1\gamma^{-1})B(\gamma,\gamma k_2\gamma^{-1})$.  Finally these two TPOs can fuse into one, shown in Figure~\ref{homke}.  Alternatively, we could have proceeded from Figure~\ref{homka} to \ref{homkd} by fusing the TPOs before shrinking the TDL.  This route takes us from Figure~\ref{homka} to \ref{homkd} and finally to \ref{homke}, while picking up the phase $B(\gamma,\gamma k_1k_2\gamma^{-1})$ along the way.  These two paths to the same result should produce the same phase, so we find the condition
\be
\label{khomcond}
B(\gamma,\gamma k_1k_2\gamma^{-1})=B(\gamma,\gamma k_1\gamma^{-1})B(\gamma,\gamma k_2\gamma^{-1}),
\ee
which tells us that $B$ is a homomorphism in its $K$ argument.

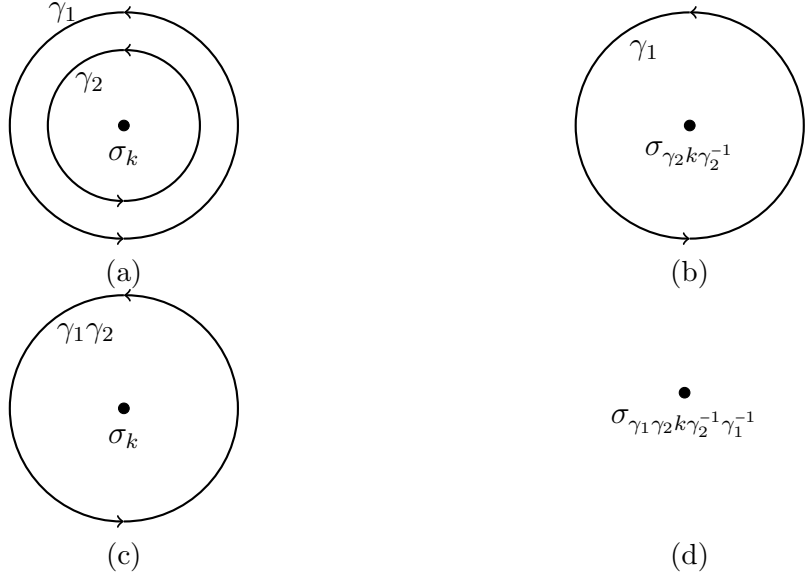
\begin{figure}
	\begin{subfigure}{0.5\textwidth}
		\centering
		\begin{tikzpicture}
		\filldraw[black] (0,0) circle (2pt);
		\node at (0,-0.4) {$\sigma_{k}$};
		\draw[thick,->] (0,0) [partial ellipse=-90:90:1cm and 1cm];
		\draw[thick,->] (0,0) [partial ellipse=90:270:1cm and 1cm];
		\node at (-0.45,0.57) {$\gamma_2$};
		\draw[thick,->] (0,0) [partial ellipse=-90:90:1.5cm and 1.5cm];
		\draw[thick,->] (0,0) [partial ellipse=90:270:1.5cm and 1.5cm];
		\node at (-0.8,1.5) {$\gamma_1$};
		\end{tikzpicture}
	\caption{}
	\label{homgama}
	\end{subfigure}
	\begin{subfigure}{0.5\textwidth}
		\centering
		\begin{tikzpicture}
		\filldraw[black] (0,0) circle (2pt);
		\node at (0,-0.4) {$\sigma_{\gamma_2 k\gamma_2^{-1}}$};
		\draw[thick,->] (0,0) [partial ellipse=-90:90:1.5cm and 1.5cm];
		\draw[thick,->] (0,0) [partial ellipse=90:270:1.5cm and 1.5cm];
		\node at (-0.6,1) {$\gamma_1$};
		\end{tikzpicture}
	\caption{}
	\label{homgamb}
	\end{subfigure}
	\begin{subfigure}{0.5\textwidth}
		\centering
		\begin{tikzpicture}
		\filldraw[black] (0,0) circle (2pt);
		\node at (0,-0.4) {$\sigma_{k}$};
		\draw[thick,->] (0,0) [partial ellipse=-90:90:1.5cm and 1.5cm];
		\draw[thick,->] (0,0) [partial ellipse=90:270:1.5cm and 1.5cm];
		\node at (-0.5,1) {$\gamma_1\gamma_2$};
		\end{tikzpicture}
	\caption{}
	\label{homgamc}
	\end{subfigure}
	\begin{subfigure}{0.5\textwidth}
		\centering
		\vspace{1cm}
		\begin{tikzpicture}
		\filldraw[black] (0,0) circle (2pt);
		\node at (0,-0.4) {$\sigma_{\gamma_1\gamma_2k\gamma_2^{-1}\gamma_1^{-1}}$};
		\end{tikzpicture}
		\vspace{1cm}
	\caption{}
	\label{homgamd}
	\end{subfigure}
	\caption{The fusion of two TDLs acting on a TPO constrains the dependence of $B$ on its $\Gamma$ argument.}
	\label{fig:homgam}
\end{figure}

We can constrain $B$'s behavior in $\Gamma$ similarly, by examining two TDLs $\gamma_1$ and $\gamma_2$ acting on a single TPO $\sigma_k$, shown in Figure~\ref{homgama}.  Shrinking the $\gamma_2$ line to act on $\sigma_k$ brings us to Figure~\ref{homgamb} and introduces a phase $B(\gamma_2,\gamma_2 k\gamma_2^{-1})$.  Then we could act on the resulting TPO with the remaining $\gamma_1$ line to arrive at Figure~\ref{homgamd}, and gain the phase $B(\gamma_1,\gamma_1\gamma_2k\gamma_2^{-1}\gamma_1^{-1})$.  Alternatively, in Figure~\ref{homgama} we could have fused the TDLs to arrive at Figure~\ref{homgamc}.  Then, acting on $\sigma_k$ with the $\gamma_1\gamma_2$ TDL would bring us to Figure~\ref{homgamd} while introducing a phase $B(\gamma_1\gamma_2,\gamma_1\gamma_2k\gamma_2^{-1}\gamma_1^{-1})$.  Again, the requirement that these two ways of reaching Figure~\ref{homgamd} produce the same phase leads to the condition
\be
B(\gamma_1\gamma_2,\gamma_1\gamma_2k\gamma_2^{-1}\gamma_1^{-1})=B(\gamma_1,\gamma_1\gamma_2k\gamma_2^{-1}\gamma_1^{-1})B(\gamma_2,\gamma_2k\gamma_2^{-1}).
\ee
Noting that an action of $\Gamma$ on $K$ taking $k\to\gamma k\gamma^{-1}$ induces an action $(\gamma^{-1}\cdot\varphi)(k)=\varphi(\gamma k\gamma^{-1})$ on $\varphi:K\to U(1)$, we can act on both sides with $\gamma_1\gamma_2$ to write the above as
\be
\label{xhomcond}
B(\gamma_1\gamma_2,k)=B(\gamma_1,k)(\gamma_1\cdot B)(\gamma_2,k),
\ee
which is the condition for $B$ to be a crossed homomorphism in $\Gamma$ and $H^1(K,U(1))$.  Considering the properties (\ref{khomcond}) and (\ref{xhomcond}) of $B$, we can regard mixed symmetries as cocycles valued in $Z^1(\Gamma,H^1(K,U(1)))$.  The following section will confirm that physical effects of these anomalies are only sensitive to the class of this cocycle in $H^1(\Gamma,H^1(K,U(1)))$ (in general with non-trivial action on the coefficients), agreeing with the results of \cite{Benini:2018reh,Yu:2020twi}.

Here we have examined mixed anomaly phases for freestanding TPOs.  Since many of the TPOs we will work with are bound to lines and cannot stand freely, for completeness appendix~\ref{sec:boundb} repeats these consistency arguments for bound TPOs.  The results match those of the freestanding case.

\subsection{Application to Orbifolds}
\label{sec:orbifolds}

We would like to take the orbifold by the zero-form symmetry $\Gamma=K.G$, where as before $K$ acts trivially (for simplicity we do not turn on any discrete torsion in $\Gamma$).  We know to expect this orbifold to exhibit decomposition, which is to say that the result should be a direct sum of orbifolds by $G$ or its subgroups.  Each non-vanishing contribution $Z_{\gamma_1,\gamma_2}$ (where $[\gamma_1,\gamma_2]=1$) to the torus partition function should contribute a term $Z_{g_1,g_2}$ to the decomposed partition function, where $g_i = \pi(\gamma_i)$ and
$\pi: \Gamma \rightarrow G$ is the projection.  In the absence of any phases, we could determine the coefficient of $Z_{g_1,g_2}$ by simply counting how many commuting pairs $\gamma_1, \gamma_2$ exist in $\Gamma$.

However, it may be that two sectors $Z_{\gamma_1, \gamma_2}$ 
and $Z_{\gamma'_1,\gamma'_2}$, for 
\begin{equation}
\pi(\gamma_1) \: = \: g_1 \: = \: \pi(\gamma'_1),
\: \: \:
\pi(\gamma_2) \: = \: g_2 \: = \: \pi(\gamma'_2),
\end{equation}
which both `project' to $Z_{g_1,g_2}$ in the $G$ orbifold, 
are related by phases.  The framework developed so far in this section will allow us to determine the contribution of such phases to the decomposed partition function.  In order to determine the coefficient of $Z_{g_1,g_2}$, the strategy will be to fix a commuting pair $\gamma_1, \gamma_2$ of elements in $\Gamma$, and use the action of the TPOs labeled by $K$ to map each $Z_{\gamma_1', \gamma_2'}$ to $Z_{\gamma_1,\gamma_2}$ (possibly accruing phases along the way).  Note that since elements of $\Gamma$ which project to the same element in $G$ differ by an element of $K$, we can set $\gamma'_1=k_1\gamma_1$ and $\gamma'_2=k_2\gamma_2$.  Once all of the contributions take this form, the sum of their coefficients should be the coefficient of $Z_{g_1,g_2}$ in the decomposed theory.  Naturally one might worry that the results of this procedure would depend on the choice of $\gamma_1$ and $\gamma_2$ -- we will return to this concern shortly.

Figure~\ref{fig:pt} illustrates this procedure.  We begin in Figure~\ref{fig:pta} by representing $Z_{k_1\gamma_1,k_2\gamma_2}$ as two TDLs labeled by $k_1\gamma_1$ and $k_2\gamma_2$ wrapping the cycles of a torus, 
pictured as its fundamental domain in $\C$.  
In Figure~\ref{fig:ptb} we insert an identity TPO on the $k_2\gamma_2$ line.  
Moving to Figure~\ref{fig:ptc}, 
we break the identity operator into an inverse pair $\sigma_{k_2}$ and $\sigma_{k_2^{-1}}$.  
This is chosen such that the line connecting them is labeled by $\gamma_2$ alone.  
Moving to Figure~\ref{fig:ptd}, we drag $\sigma_{k_2}$ across the $k_1\gamma_1$ line.  
This will effectively act on the TPO to produce $\sigma_{k_1\gamma_1k_2\gamma_1^{-1}k_1^{-1}}$, 
with the incoming line labeled by $k_1\gamma_1k_2^{-1}\gamma_1^{-1}k_1^{-1}k_2\gamma_2$.  
Moving from \ref{fig:ptc} to \ref{fig:ptd} 
will introduce the phase $B(k_1\gamma_1,k_1\gamma_1k_2\gamma_1^{-1}k_1^{-1})=(g_1^{-1}\cdot B)(k_1\gamma_1,k_2)$ due to a potential mixed anomaly.

\begin{figure}
	\begin{subfigure}{0.5\textwidth}
		\centering
		\begin{tikzpicture}
		\draw[thin] (0,0)--(5,0);
		\draw[thin] (5,0)--(5,5);
		\draw[thin] (0,0)--(0,5);
		\draw[thin] (0,5)--(5,5);
		\draw[very thick,->] (1.5,0)--(2,1.25);
		\draw[very thick] (2,1.25)--(2.5,2.5);
		\draw[very thick,->] (2.5,2.5)--(3,3.75);
		\draw[very thick] (3,3.75)--(3.5,5);
		\draw[very thick] (4.25,2.5)--(5,2.5);
		\draw[very thick,->] (2.5,2.5)--(4.25,2.5);
		\draw[very thick] (2.5,2.5)--(1.5,2.5);
		\draw[very thick] (0.75,2.5)--(1.5,2.5);
		\draw[very thick,->] (0,2.5)--(0.75,2.5);
		\node at (2.5,4) {$k_1\gamma_1$};
		\node at (2.5,1) {$k_1\gamma_1$};
		\node at (1.2,2.8) {$k_2\gamma_2$};
		\node at (3.8,2.8) {$k_2\gamma_2$};
		\end{tikzpicture}
		\caption{}
		\label{fig:pta}
	\end{subfigure}
	\begin{subfigure}{0.5\textwidth}
		\centering
		\begin{tikzpicture}
		\draw[thin] (0,0)--(5,0);
		\draw[thin] (5,0)--(5,5);
		\draw[thin] (0,0)--(0,5);
		\draw[thin] (0,5)--(5,5);
		\draw[very thick,->] (1.5,0)--(2,1.25);
		\draw[very thick] (2,1.25)--(2.5,2.5);
		\draw[very thick,->] (2.5,2.5)--(3,3.75);
		\draw[very thick] (3,3.75)--(3.5,5);
		\draw[very thick] (4.25,2.5)--(5,2.5);
		\draw[very thick,->] (2.5,2.5)--(4.25,2.5);
		\draw[very thick] (2.5,2.5)--(1.5,2.5);
		\draw[very thick] (0.75,2.5)--(1.5,2.5);
		\draw[very thick,->] (0,2.5)--(0.75,2.5);
		\node at (2.5,4) {$k_1\gamma_1$};
		\node at (2.5,1) {$k_1\gamma_1$};
		\node at (1,2.8) {$k_2\gamma_2$};
		\node at (4.5,2.8) {$k_2\gamma_2$};
		\filldraw[black] (1.5,2.5) circle (2pt);
		\node at (1.5,2.2) {$\sigma_1$};
		\end{tikzpicture}
		\caption{}
		\label{fig:ptb}
	\end{subfigure}
	\begin{subfigure}{0.5\textwidth}
		\centering
		\begin{tikzpicture}
		\draw[thin] (0,0)--(5,0);
		\draw[thin] (5,0)--(5,5);
		\draw[thin] (0,0)--(0,5);
		\draw[thin] (0,5)--(5,5);
		\draw[very thick,->] (1.5,0)--(2,1.25);
		\draw[very thick] (2,1.25)--(2.5,2.5);
		\draw[very thick,->] (2.5,2.5)--(3,3.75);
		\draw[very thick] (3,3.75)--(3.5,5);
		\draw[very thick] (3.75,2.5)--(5,2.5);
		\draw[very thick,->] (2.5,2.5)--(3.75,2.5);
		\draw[very thick] (2.5,2.5)--(1.5,2.5);
		\draw[very thick,->] (0,2.5)--(1.5,2.5);
		\node at (2.5,4) {$k_1\gamma_1$};
		\node at (2.5,1) {$k_1\gamma_1$};
		\node at (1.25,2.8) {$\gamma_2$};
		\node at (3.75,2.8) {$k_2\gamma_2$};
		\filldraw[black] (2,2.5) circle (2pt);
		\node at (2,2.2) {$\sigma_{k_2}$};
		\filldraw[black] (0.5,2.5) circle (2pt);
		\node at (0.5,2.2) {$\sigma_{k_2^{-1}}$};
		\end{tikzpicture}
		\caption{}
		\label{fig:ptc}
	\end{subfigure}
	\begin{subfigure}{0.5\textwidth}
		\centering
		\begin{tikzpicture}
		\draw[thin] (0,0)--(5,0);
		\draw[thin] (5,0)--(5,5);
		\draw[thin] (0,0)--(0,5);
		\draw[thin] (0,5)--(5,5);
		\draw[very thick,->] (1.5,0)--(2,1.25);
		\draw[very thick] (2,1.25)--(2.5,2.5);
		\draw[very thick,->] (2.5,2.5)--(3,3.75);
		\draw[very thick] (3,3.75)--(3.5,5);
		\draw[very thick] (4.5,2.5)--(5,2.5);
		\draw[very thick,->] (2.5,2.5)--(4.5,2.5);
		\draw[very thick] (2.5,2.5)--(1.5,2.5);
		\draw[very thick] (0.75,2.5)--(1.5,2.5);
		\draw[very thick,->] (0,2.5)--(0.75,2.5);
		\node at (2.5,4) {$k_1\gamma_1$};
		\node at (2.5,1) {$k_1\gamma_1$};
		\node[scale=1] at (2,2.8) {$\gamma_2$};
		\node at (0.5,2.8) {$k_2\gamma_2$};
		\node at (4.5,2.8) {$k_2\gamma_2$};
		\filldraw[black] (1.5,2.5) circle (2pt);
		\node at (1.5,2.2) {$\sigma_{k_2^{-1}}$};
		\filldraw[black] (3.0,2.5) circle (2pt);
		\node at (3.75,2.2) {$\sigma_{k_1\gamma_1k_2\gamma_1^{-1}k_1^{-1}}$};
		\end{tikzpicture}
		\caption{}
		\label{fig:ptd}
	\end{subfigure}
	\begin{subfigure}{0.5\textwidth}
		\centering
		\begin{tikzpicture}
		\draw[thin] (0,0)--(5,0);
		\draw[thin] (5,0)--(5,5);
		\draw[thin] (0,0)--(0,5);
		\draw[thin] (0,5)--(5,5);
		\draw[very thick,->] (1.5,0)--(2,1.25);
		\draw[very thick] (2,1.25)--(2.5,2.5);
		\draw[very thick,->] (2.5,2.5)--(3,3.75);
		\draw[very thick] (3,3.75)--(3.5,5);
		\draw[very thick] (4.0,2.5)--(5,2.5);
		\draw[very thick,->] (2.5,2.5)--(4.0,2.5);
		\draw[very thick] (2.5,2.5)--(1.5,2.5);
		\draw[very thick,->] (0,2.5)--(1.5,2.5);
		\node at (2.5,4) {$k_1\gamma_1$};
		\node at (2.5,1) {$k_1\gamma_1$};
		\node at (1.25,2.8) {$k_2\gamma_2$};
		\node at (3.5,2.2) {$\gamma_2$};
		\filldraw[black] (2.0,2.5) circle (2pt);
		\node at (2.0,2.2) {$\sigma_{k_2^{-1}}$};
		\filldraw[black] (4.5,2.5) circle (2pt);
		\node at (4.5,2.2) {$\sigma_{k_2}$};
		\filldraw[black] (2.5,2.5) circle (2pt);
		\node[scale=1.0,rotate=00] at (3.75,2.75) {$\sigma_{[k_2^{-1},k_1\gamma_1]}$};
		\end{tikzpicture}
		\caption{}
		\label{fig:pte}
	\end{subfigure}
	\begin{subfigure}{0.5\textwidth}
		\centering
		\begin{tikzpicture}
		\draw[thin] (0,0)--(5,0);
		\draw[thin] (5,0)--(5,5);
		\draw[thin] (0,0)--(0,5);
		\draw[thin] (0,5)--(5,5);
		\draw[very thick,->] (1.5,0)--(2,1.25);
		\draw[very thick] (2,1.25)--(2.5,2.5);
		\draw[very thick,->] (2.5,2.5)--(3,3.75);
		\draw[very thick] (3,3.75)--(3.5,5);
		\draw[very thick] (3.75,2.5)--(5,2.5);
		\draw[very thick,->] (2.5,2.5)--(3.75,2.5);
		\draw[very thick] (2.5,2.5)--(1.5,2.5);
		\draw[very thick,->] (0,2.5)--(1.5,2.5);
		\node at (2.5,4) {$k_1\gamma_1$};
		\node at (2.5,1) {$k_1\gamma_1$};
		\node at (1.25,2.8) {$k_2\gamma_2 $};
		\node at (3.75,2.2) {$\gamma_2$};
		\filldraw[black] (2.0,2.5) circle (2pt);
		\node at (2.0,2.2) {$\sigma_{k_2^{-1}}$};
		\filldraw[black] (0.5,2.5) circle (2pt);
		\node at (0.5,2.2) {$\sigma_{k_2}$};
		\filldraw[black] (2.5,2.5) circle (2pt);
		\node[scale=1.0,rotate=00] at (3.75,2.75) {$\sigma_{[k_2^{-1},k_1\gamma_1]}$};
		\end{tikzpicture}
		\caption{}
		\label{fig:ptf}
	\end{subfigure}
	\begin{subfigure}{0.5\textwidth}
		\centering
		\begin{tikzpicture}
		\draw[thin] (0,0)--(5,0);
		\draw[thin] (5,0)--(5,5);
		\draw[thin] (0,0)--(0,5);
		\draw[thin] (0,5)--(5,5);
		\draw[very thick,->] (1.5,0)--(2,1.25);
		\draw[very thick] (2,1.25)--(2.5,2.5);
		\draw[very thick,->] (2.5,2.5)--(3,3.75);
		\draw[very thick] (3,3.75)--(3.5,5);
		\draw[very thick] (4.25,2.5)--(5,2.5);
		\draw[very thick,->] (2.5,2.5)--(4.25,2.5);
		\draw[very thick] (2.5,2.5)--(1.5,2.5);
		\draw[very thick,->] (0,2.5)--(1.5,2.5);
		\node at (2.5,4) {$k_1\gamma_1$};
		\node at (2.5,1) {$k_1\gamma_1$};
		\node at (1.5,2.8) {$\gamma_2$};
		\node at (4.25,2.2) {$\gamma_2$};
		\filldraw[black] (0.75,2.5) circle (2pt);
		\node at (0.75,2.2) {$\sigma_1$};
		\filldraw[black] (2.5,2.5) circle (2pt);
		\node[scale=1.0,rotate=00] at (3.75,2.75) {$\sigma_{[k_2^{-1},k_1\gamma_1]}$};
		\end{tikzpicture}
		\caption{}
		\label{fig:ptg}
	\end{subfigure}
	\begin{subfigure}{0.5\textwidth}
		\centering
		\begin{tikzpicture}
		\draw[thin] (0,0)--(5,0);
		\draw[thin] (5,0)--(5,5);
		\draw[thin] (0,0)--(0,5);
		\draw[thin] (0,5)--(5,5);
		\draw[very thick,->] (1.5,0)--(2,1.25);
		\draw[very thick] (2,1.25)--(2.5,2.5);
		\draw[very thick,->] (2.5,2.5)--(3,3.75);
		\draw[very thick] (3,3.75)--(3.5,5);
		\draw[very thick] (4.25,2.5)--(5,2.5);
		\draw[very thick,->] (2.5,2.5)--(4.25,2.5);
		\draw[very thick] (2.5,2.5)--(1.5,2.5);
		\draw[very thick,->] (0,2.5)--(1.5,2.5);
		\node at (2.5,4) {$k_1\gamma_1$};
		\node at (2.5,1) {$k_1\gamma_1$};
		\node at (1.5,2.8) {$\gamma_2$};
		\node at (4.25,2.2) {$\gamma_2$};
		\filldraw[black] (2.5,2.5) circle (2pt);
		\node[scale=1.0,rotate=00] at (3.75,2.75) {$\sigma_{[k_2^{-1},k_1\gamma_1]}$};
		\end{tikzpicture}
		\caption{}
		\label{fig:pth}
	\end{subfigure}
	\caption{We map the $k_1\gamma_1, k_2\gamma_2$ sector of the orbifold partition function into the $k_1\gamma_1,\gamma_2$ sector.}
	\label{fig:pt}
\end{figure}

In anticipation of the upcoming steps, we split $\sigma_{k_1\gamma_1k_2\gamma_1^{-1}k_1^{-1}}$ into $\sigma_{[k_2^{-1},k_1\gamma_1]}$ and $\sigma_{k_2}$.  The result of this is shown in Figure~\ref{fig:pte}, where the TPO $\sigma_{[k_2^{-1},k_1\gamma_1]}$ has been moved to the intersection.  This allows the horizontal lines on both sides of the intersection to be labeled by $\gamma_2$, and we have effectively moved $\sigma_{k_2}$ across the junction.  
In Figure~\ref{fig:ptf} we approach $\sigma_{k_2^{-1}}$ with its inverse from the other side, joining them once again into an identity operator.  This leaves us with $Z_{k_1\gamma_1, \gamma_2}$ with the point operator $\sigma_{[k_2^{-1},k_1\gamma_1]}$ sitting at the junction.  The reason for this extra insertion is that, 
while by assumption $[\gamma_1,\gamma_2]=[k_1\gamma_1,k_2\gamma_2]=1$, it is not necessarily true that $[k_1\gamma_1,\gamma_2]=1$.

This is not an issue, as the next move would be to repeat the process and map $Z_{k_1\gamma_1,\gamma_2}$ to $Z_{\gamma_1,\gamma_2}$.  The steps are essentially the same, so we will not illustrate this part separately.  The important fact is that in doing so, we would generate the phase $B^{-1}(\gamma_2,\gamma_2k_1\gamma_2^{-1})=[(g_2^{-1}\cdot B)(\gamma_2,k_1)]^{-1}$.  The inverse appears because we must either move $\sigma_{k_1}$ across $\gamma_2$ with orientation opposite to $k_1\gamma_1$ in the previous steps, or equivalently drag $\sigma_{k_1^{-1}}$ across with the same orientation.  In moving $\sigma_{k_1}$ across $\gamma_2$, we generate a point operator which is inverse to $\sigma_{[k_2^{-1},k_1\gamma_1]}$, such that the resulting diagram for $Z_{\gamma_1,\gamma_2}$ contains no more non-identity TPOs (as is consistent with the group law, since $\gamma_1$ and $\gamma_2$ commute).

Of course, the order in which we perform this reduction -- $Z_{k_1\gamma_1,k_2\gamma_2}\to Z_{k_1\gamma_1, \gamma_2}\to Z_{\gamma_1,\gamma_2}$ versus $Z_{k_1\gamma_1,k_2\gamma_2}\to Z_{\gamma_1, k_2\gamma_2}
\to Z_{\gamma_1, \gamma_2}$ -- should not matter.\footnote{More specifically, because all of these steps are invertible, we could take one path and then the inverse of the other to map $Z_{k_1\gamma_1,k_2\gamma_2}$ to itself.  Because the phase we assign to a given sector should be unambiguous, the total phase accrued this way should be unity.  Therefore the two ways of reaching $Z_{\gamma_1,\gamma_2}$ should produce the same phase.}  The order would seem to affect the $\Gamma$ argument of $B$.  In order to guarantee that these two paths to $Z_{\gamma_1,\gamma_2}$ are equivalent, we take $B$ to be a pullback from an element of $H^1(G,H^1(K,U(1)))$.  
Heuristically, this means that we are free to modify how the effective part of the symmetry acts on the TPOs, but not the non-effective part.

In summary, we have used manipulations with TPOs to derive a relationship between different contributions to a $\Gamma$ orbifold partition function with trivially-acting normal subgroup $K$:\footnote{It may be helpful to note that, for $\Gamma=K.G$, the action of $G$ on $\varphi\in H^1(K,U(1))$ can be written as $(g\cdot\varphi)(k)=\varphi(\gamma^{-1}k\gamma)$ for any $\gamma$ satisfying $\pi(\gamma)=g$.  In particular, this makes it clear that such an action does not depend on any choice of section for $\Gamma$.}
\begin{equation}
\label{mixedanomphases}
Z_{k_1\gamma_1, k_2\gamma_2} \: = \:
 \frac{(g_1^{-1}\cdot B)(g_1,k_2)
}{
(g_2^{-1}\cdot B)(g_2,k_1)} 
\, Z_{\gamma_1, \gamma_2},
\end{equation}
where $g_i=\pi(\gamma_i)$, with $\pi$ the projection map $\Gamma\to G$ appearing in the extension (\ref{gammaext}).  In \cite{Robbins:2021lry,Robbins:2021ibx,Robbins:2021xce}, where these mixed anomaly phases were studied under the name `quantum symmetry phases,' a special version of this relation was taken as an ansatz.  Here we see how it follows naturally from the topological operator formulation of trivial symmetries.  In the following section we check that these phases produce reasonable results for the decomposed orbifold theory.

\subsection{Consistency Checks}

Once each sector with fixed $g_1=\pi(\gamma_1), g_2=\pi(\gamma_2)$ is mapped to a phase times $Z_{\gamma_1,\gamma_2}$, we expect that the coefficient of $Z_{g_1,g_2}$ appearing in the decomposed partition function should be given by the sum of those phases.  Explicitly, this is
\be
\label{f1}
F(\gamma_1,\gamma_2) \: = \:
\sum_{\substack{k_1,k_2 \in K \\ [k_1\gamma_1, k_2\gamma_2] = 1}}
\frac{(g_1^{-1}\cdot B)(g_1,k_2)}{(g_2^{-1}\cdot B)(g_2,k_1)}
\ee
for a commuting pair $\gamma_1,\gamma_2$.

\subsubsection{Reference (In)dependence}
\label{sec:refind}

Now we return to the question of dependence on $\gamma_1,\gamma_2$.  In some cases, there will be a canonical choice of $\gamma_1,\gamma_2$.  Consider a split extension, i.e.~the extension class is trivial.  Then there is always a choice of section such that $[(1,g_1),(1,g_2)]=1$.  In such a case, the relative phases that arise from $B$ when mapping each sector in the $\Gamma$ orbifold to its corresponding $Z_{(1,g_1),(1,g_2)}$ are in fact redundant with phases arising from a choice of discrete torsion in the $\Gamma$ orbifold.\footnote{The precise condition for mixed anomaly phases to be redundant with discrete torsion is that the cup product of $B\in H^1(G,H^1(K,U(1)))$ and the extension class $c\in H^2(G,K)$ must be trivial in $H^3(G,U(1))$ \cite[Section 2.3]{Robbins:2021ibx}.  This is manifestly satisfied for split extensions.}  There will be other cases, however, where there is no obvious choice of reference, and we must be more careful in treating the decomposition process.

In order to see how coefficients in the decomposed partition function depend on $\gamma_1,\gamma_2$, suppose we had picked a different commuting pair, $\tilde{\gamma}_1,\tilde{\gamma}_2$ (with the same $G$ projection).  Then we would have found the coefficient of $Z_{g_1,g_2}$ to be
\be
\label{f2}
F(\tilde{\gamma}_1,\tilde{\gamma}_2) \: = \:
\sum_{\substack{\tilde{k}_1,\tilde{k}_2 \in K \\ [\tilde{k}_1\tilde{\gamma}_1,\tilde{k}_2\tilde{\gamma}_2]=1}}
\frac{(g_1^{-1}\cdot B)(g_1,\tilde{k}_2)}{(g_2^{-1}\cdot B)(g_2,\tilde{k}_1)}.
\ee
Since $\pi(\gamma_1)=\pi(\tilde{\gamma}_1)=g_1$ and $\pi(\gamma_2)=\pi(\tilde{\gamma_2})=g_2$, we can find $\k_1$ and $\k_2$ such that $\tilde{\gamma}_1=\k_1\gamma_1$ and $\tilde{\gamma}_2=\k_2\gamma_2$.  By redefining our summation variable to $k_i=\tilde{k}_i\k_i$ in (\ref{f2}), we see that these two expressions relate to each other as
\be
\label{refshiftcond}
F(\gamma_1,\gamma_2)=F(\tilde{\gamma}_1,\tilde{\gamma}_2)\frac{(g_2^{-1}\cdot B)(g_2,\k_1)}{(g_1^{-1}\cdot B)(g_1,\k_2)},
\ee
which is to say that choosing a different `reference' sector in the $\Gamma$ orbifold would change the computed coefficient for $Z_{g_1,g_2}$ by a phase -- for consistency, such a choice should not affect the result.

There are two ways in which $F(\gamma_1,\gamma_2)$ could depend only on $g_1$ and $g_2$.  It could be that the summand of (\ref{f1}) is identically trivial, in which case the value of $F$ counts the number of commuting pairs in $\Gamma$ with fixed $G$ projection; this number is clearly independent of any choice of reference.  Alternatively, if $B$ provides any non-trivial phases to the sum, it should be that those phases sum to zero.  From (\ref{refshiftcond}) we see that because changing our reference sector multiplies the entire sum by a phase, $F(\gamma_1,\gamma_2)=0$ would also be a reference-independent statement.

Consider the case where $K$ is central in $\Gamma$.  Then the condition $[k_1\gamma_1,k_2\gamma_2]=1$ is equivalent to $[\gamma_1,\gamma_2]=1$, which $\gamma_1$ and $\gamma_2$ were specifically chosen to satisfy.  This means that the sum in (\ref{f1}) is unconstrained and runs over two full copies of $K$.  If we fix an element $k_1$, the sum over $k_2$ is a sum over a homomorphism from $K$ into $U(1)$, so either the homomorphism is trivial or the sum vanishes.  The same is true for any value of $k_1$, which means that for central $K$ we would calculate the same coefficient no matter which sector was chosen as the reference point.

Unfortunately, such reference independence may not be present for non-central $K$.  Consider, for example, $\Gamma=S_3\times\Z_2$ with $K=S_3$.  We see from 
\be
H^1(\Z_2,H^1(S_3,U(1)))=H^1(\Z_2,\Z_2)=\Z_2
\ee 
that there is the possibility of a non-trivial mixed anomaly in this example.  With this mixed anomaly present, one can apply decomposition to express the $\Gamma$ orbifold sectors in terms of $\Z_2$ orbifold sectors as
\be
\frac{1}{12}[18Z_{0,0}+6(a_1Z_{0,1}+a_2Z_{1,0}+a_3Z_{1,1})]
\ee
where choice of reference sector allows for any choice of the coefficients $a_i$ satisfying $a_1^2=a_2^2=a_3^2=1$.  As discussed above, because this is a split extension, we have the canonical choice of using $Z_{(1,g_1),(1,g_2)}$ as the reference point in each sector.  Doing so, we get $a_1=a_2=a_3=1$, which indeed is the only choice that corresponds to a sensible CFT partition function -- the decomposed orbifold is then equivalent to one copy of the parent theory plus one copy of the orbifold by the effectively-acting $\Z_2$.  This choice of reference is also the one which for which the mixed anomaly phases are redundant with discrete torsion in $\Gamma$.

In more general situations where the extension is neither central nor split, we do not have a general proof of consistency of this procedure, so we would need to verify case by case that choice of reference sector does not lead to any issues.  In the examples presented in the following section, only the one examined in section~\ref{sec:q8ex} is neither split nor central.  However, in calculating $F$ for that example, $k_1$ and $k_2$ will always fill out a subgroup of $K^2$; this means that any non-trivial contributions to the sum will cause it to vanish, as in the central case.

\subsubsection{Coboundary Invariance}

Next, we can check that $B$ that are exact in $H^1(G,H^1(K,U(1)))$ do not contribute.  Such a cocycle takes the form
\be
\label{exactb}
B(g,k) \: = \: \frac{(g\cdot \varphi)(k)}{\varphi(k)}
\ee
for $\varphi\in H^1(K,U(1))$.  With such a choice of $B$, 
the relative phase (\ref{mixedanomphases}) between $Z_{k_1\gamma_1, k_2\gamma_2}$ and 
$Z_{\gamma_1, \gamma_2}$ becomes
\be
\label{exactphase1}
\frac{\varphi(k_2)\varphi(\gamma_2k_1\gamma_2^{-1})}{\varphi(k_1)\varphi(\gamma_1k_2\gamma_1^{-1})}.
\ee
Recall that the values of $k$ are such that $\gamma'_1=k_1\gamma_1$ and $\gamma'_2=k_2\gamma_2$ is a commuting pair.  Expressing (\ref{exactphase1}) in terms of these variables and collecting terms produces
\be
\varphi(\gamma'_2\gamma'_1\gamma_1^{-1}\gamma_2^{-1}\gamma_1\gamma_2{\gamma'_2}^{-1}{\gamma'_1}^{-1}),
\ee
which is manifestly trivial since $[\gamma_1,\gamma_2]=[\gamma'_1,\gamma'_2]=1$, so the phases appearing between sectors from $B$ are invariant under coboundary shifts.  $B$ then only matters up to its class in $H^1(G,H^1(K,U(1)))$, as claimed earlier.

\subsubsection{Modular Invariance}

Finally, we can check that coefficients calculated in this way do not disrupt modular invariance in the decomposed orbifold.  When the mixed anomaly phases are redundant with discrete torsion in $\Gamma$, we simply have decomposition with discrete torsion as studied in \cite{Robbins:2020msp}, and modular invariance should hold automatically.  It remains to verify modular invariance when the mixed anomaly phases cannot be described by discrete torsion.  Following the discussion of section~\ref{sec:refind} above, let us assume that we have already verified reference independence in the example at hand.

As a recap, we have the coefficient of $Z_{g_1,g_2}$ in the post-decomposition partition function calculated in (\ref{f1}) as
\be
F(\gamma_1,\gamma_2) \: = \:
\sum_{\substack{k_1,k_2 \in K \\ [k_1\gamma_1 , k_2\gamma_2]=1}}\frac{(g_1^{-1}\cdot B)(g_1,k_2)
}{(g_2^{-1}\cdot B)(g_2,k_1)}.
\ee
The input here is a pair $\gamma_1,\gamma_2$ of elements of $\Gamma$ satisfying 
\be
\label{gammaconst}
\pi(\gamma_1)=g_1,\hspace{0.5cm} \pi(\gamma_2)=g_2,\hspace{0.5cm} [\gamma_1,\gamma_2]=1,
\ee
though as argued above we are free to choose any such $\gamma_1,\gamma_2$ satisfying these conditions without changing the answer.  Now imagine that we apply a modular transformation to $g_1$ and $g_2$ -- call the result $\bar{g}_1$ and $\bar{g}_2$.  We would calculate the coefficient of $Z_{\bar{g}_1,\bar{g}_2}$ as
\be
F(\bar{\gamma}_1,\bar{\gamma}_2) \: = \:
\sum_{\substack{\bar{k}_1,\bar{k}_2 \in K \\ [\bar{k}_1\bar{\gamma}_1 , \bar{k}_2\bar{\gamma}_2]=1}}\frac{(\bar{g}_1^{-1}\cdot B)(\bar{g}_1,\bar{k}_2)
}{(\bar{g}_2^{-1}\cdot B)(\bar{g}_2,\bar{k}_1)}
\ee
for any $\bar{\gamma}_1,\bar{\gamma}_2$ satisfying 
\be
\label{gammabarconst}
\pi(\bar{\gamma}_1)=\bar{g}_1,\hspace{0.5cm} \pi(\bar{\gamma}_2)=\bar{g}_2,\hspace{0.5cm}[\bar{\gamma}_1,\bar{\gamma}_2]=1.
\ee

Take 
\be
\bar{g}_1=g_2,\hspace{.5cm} \bar{g}_2=g_1^{-1}.
\ee
This is a pair related by the modular $S$ transformation.  Our goal will be to relate $F(\bar{\gamma}_1,\bar{\gamma}_2)$ to $F(\gamma_1,\gamma_2)$.  It is then quite natural to make the choice
\be
\bar{\gamma}_1=\gamma_2,\hspace{.5cm}\bar{\gamma}_2=\gamma_1^{-1}.
\ee
Thanks to the stipulated properties (\ref{gammaconst}) of $\gamma_1, \gamma_2$, this choice of $\bar{\gamma}_1,\bar{\gamma}_2$ manifestly satisfies (\ref{gammabarconst}).  With these choices, $F(\bar{\gamma}_1,\bar{\gamma}_2)$ takes the form
\be
F(\bar{\gamma}_1,\bar{\gamma}_2)=F(\gamma_2,\gamma_1^{-1})=\sum_{\substack{\bar{k}_1,\bar{k}_2\in K \\ [\bar{k}_1\gamma_2,\bar{k}_2\gamma_1^{-1}]=1}}\frac{(g_2^{-1}\cdot B)(g_2,\bar{k}_2)}{(g_1\cdot B)(g_1^{-1},\bar{k}_1)}.
\ee
Of course, we are free to redefine our summation variables, at the cost of changing the commutation constraint to compensate.  We will shift to a sum over $k_1$ and $k_2$ given by
\be
\bar{k}_1=\gamma_1k_2\gamma_1^{-1}, \hspace{0.5cm} \bar{k}_2=k_1^{-1}.
\ee
With this redefinition, the commutation constraint appearing in the sum takes the form $[\gamma_1k_2\gamma_1^{-1}\gamma_2,k_1^{-1}\gamma_1^{-1}]=1$.  Writing out this constraint and keeping in mind that $\gamma_1$ commutes with $\gamma_2$, it is straightforward to rearrange it to produce $[k_1\gamma_1,k_2\gamma_2]=1$, as appearing in the original $F(\gamma_1,\gamma_2)$ sum.  Now that we are summing over the same set of $k_1,k_2$ in each version, it remains to compare the summands.  With all of the choices made so far, we have
\be
F(\gamma_2,\gamma_1^{-1})=\sum_{\substack{k_1,k_2\in K \\ [k_1\gamma_1,k_2\gamma_2]=1}}\frac{(g_2^{-1}\cdot B)(g_2,k_1^{-1})}{(g_1\cdot B)(g_1^{-1},\gamma_1k_2\gamma_1^{-1})}.
\ee
Keeping in mind the properties (\ref{khomcond}) and (\ref{xhomcond}) of $B$, one readily transforms the above summand to match the one in (\ref{f1}), giving $F(\gamma_1,\gamma_2)=F(\gamma_2,\gamma_1^{-1})$.

Now we would like to repeat this process for pairs in $G$ related by the modular $T$ transformation.  We can achieve this by choosing
\be
\bar{g}_1=g_1,\hspace{.5cm} \bar{g}_2=g_1g_2.
\ee
As before, a convenient choice of $\bar{\gamma}_1$ and $\bar{\gamma}_2$ mimics the choice of $\bar{g}$ and is given by
\be
\bar{\gamma}_1=\gamma_1,\hspace{.5cm}\bar{\gamma}_2=\gamma_1\gamma_2.
\ee
With these choices,
\be
F(\bar{\gamma}_1,\bar{\gamma}_2)=F(\gamma_1,\gamma_1\gamma_2)=\sum_{\substack{\bar{k}_1,\bar{k}_2\in K \\ [\bar{k}_1\gamma_1,\bar{k}_2\gamma_1\gamma_2]=1}}\frac{(g_1^{-1}\cdot B)(g_1,\bar{k}_2)}{[(g_1g_2)^{-1}\cdot B](g_1g_2,\bar{k}_1)}.
\ee
Shifting our $\bar{k}$ to
\be
\bar{k}_1=k_1, \hspace{0.5cm} \bar{k}_2=k_2\gamma_2k_1\gamma_2^{-1}
\ee
gives us the commutation constraint $[k_1\gamma_1,k_1\gamma_1k_2\gamma_2]=1$, again equivalent to $[k_1\gamma_1,k_2\gamma_2]=1$.  At this point we have
\be
F(\gamma_1,\gamma_1\gamma_2)=\sum_{\substack{k_1,k_2\in K \\ [k_1\gamma_1,k_2\gamma_2]=1}}\frac{(g_1^{-1}\cdot B)(g_1,k_2\gamma_2k_1\gamma_2^{-1})}{[(g_1g_2)^{-1}\cdot B](g_1g_2,k_1)}.
\ee
Manipulations similar to those used in the previous case restore the initial form of the summand, leading to $F(\gamma_1,\gamma_1\gamma_2)=F(\gamma_1,\gamma_2)$.

\subsubsection{Summary}

To recap the meaning of these many calculations, when we apply decomposition to express the genus one partition function of an orbifold by $\Gamma$ with trivially-acting subgroup $K$ in terms of a direct sum of orbifolds by subgroups of $G=\Gamma/K$, a mixed anomaly between the $K$ TPOs and the $\Gamma$ TDLs can contribute relative phases between the sectors of the $\Gamma$ orbifold.  These phases are given explicitly in (\ref{mixedanomphases}), and depend only on the cohomology class of $B$ in $H^1(G,H^1(K,U(1)))$ pulled back to $H^1(\Gamma,H^1(K,U(1)))$.  These phases are sometimes, but not always, redundant with discrete torsion in $\Gamma$.  The coefficient of $Z_{g_1,g_2}$ in the decomposed orbifold is given by the sum of these phases as (\ref{f1}) -- this function $F$ gives the coefficient of $Z_{\pi(\gamma_1),\pi(\gamma_2)}=Z_{g_1,g_2}$ for a fixed commuting pair $\gamma_1,\gamma_2$.  Sometimes there will be a canonical choice of $\gamma_1,\gamma_2$; when there is not, one may need to verify that the results will not depend on such a choice.  The coefficients obtained this way are invariant under the generators of the modular group, and therefore modular invariance is not compromised by the phases arising from $B$.  This formulation not only extends the results of studying such phases in \cite{Robbins:2021lry,Robbins:2021ibx,Robbins:2021xce}, it provides physical insight to their origin as mixed anomalies.

\subsection{Mixed Gauging}

So far we have discussed gauging a zero-form symmetry $\Gamma$ with trivially-acting subgroup $K$.  
In general, the resulting theory has multiple vacua, and a one-form symmetry.  
Gauging that one-form symmetry projects the theory onto the component in the direct sum associated with that vacuum, as discussed in \cite{Sharpe:2019ddn}.  In general, we expect subsequent gaugings of a theory to be composable, so this operation should be expressible as a gauging of the original theory.  In fact, we can give a relatively simple prescription for the result of such a gauging, motivated by a condensation defect construction in \cite[section 3]{Lin:2022xod}.

For clarity, let us refer schematically to the original theory as $T$.  Gauging the zero-form symmetry $\Gamma_{[0]}$
yields a theory denoted $[T/\Gamma_{[0]}]$, and gauging the one-form symmetry\footnote{In the mathematics literature, one-form symmetries have been denoted $BK$ for several decades; here, we use a more recent convention to denote them by $K_{[1]}$.} $K_{[1]}$ of the resulting theory brings us to $[ [T/\Gamma_{[0]}]/K_{[1]}]$.  A correlation function in $[[T/\Gamma_{[0]}]/K_{[1]}]$ can be thought of as a correlation function in $[T/\Gamma_{[0]}]$ with a projector $\Pi_R$ inserted, given by
\begin{equation}
\label{mg_pi}
\Pi_{R} \: = \: \sum_i \frac{ \dim R_i }{ |K| }
\sum_{k \in K} \chi_{R_i}(k^{-1}) \, \sigma_k,
\end{equation}
following  (\ref{eq:proj}) and (\ref{projformula}).  Here each $R_i$ is an irreducible representation, and we sum over orbits of irreducible representations related by the action of $G$ on $K$.

\begin{figure}
	\centering
	\begin{tikzpicture}
	\draw[thin] (0,0)--(5,0);
	\draw[thin] (5,0)--(5,5);
	\draw[thin] (0,0)--(0,5);
	\draw[thin] (0,5)--(5,5);
	\draw[very thick,->] (1.5,0)--(2,1.25);
	\draw[very thick] (2,1.25)--(2.5,2.5);
	\draw[very thick,->] (2.5,2.5)--(3,3.75);
	\draw[very thick] (3,3.75)--(3.5,5);
	\draw[very thick] (4.25,2.5)--(5,2.5);
	\draw[very thick,->] (2.5,2.5)--(4.25,2.5);
	\draw[very thick] (2.5,2.5)--(1.5,2.5);
	\draw[very thick] (0.75,2.5)--(1.5,2.5);
	\draw[very thick,->] (0,2.5)--(0.75,2.5);
	\node at (2.5,4) {$\gamma_1$};
	\node at (2.5,1) {$\gamma_1$};
	\node at (1,2.8) {$\gamma_2$};
	\node at (4.5,2.8) {$\gamma_2$};
	\filldraw[black] (2.5,2.5) circle (2pt);
	\node at (2.75,2.25) {$\sigma_k$};
	\end{tikzpicture}
	\caption{The visual representation of $Z_{\gamma_1,\gamma_2,\sigma_k}$.}
	\label{fig:mg}
\end{figure}
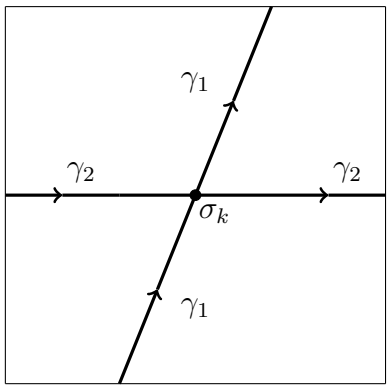

A correlation function of $[T/\Gamma_{[0]}]$ with $\Pi_R$ inserted is in turn calculated as a sum (with coefficents from (\ref{mg_pi})) of correlation functions of $T$, each with an insertion of the TPO $\sigma_k$.  Applying this logic to the genus one partition function, we expect the $[ [T/\Gamma_{[0]}]/K_{[1]}]$ partition function to be expressible as a sum of the form\footnote{Schematically, we might regard this as a gauging of $T$ by an extension of a one-form symmetry by a zero-form symmetry.  Appendix~\ref{app:mixedext} provides some additional insight to the interpretation of the symmetry of the ungauged theory as such a mixed extension.}
\be
\label{mg_pf}
\sum_i\frac{\text{dim}R_i}{|K||\Gamma|}\sum_{\substack{\gamma_1,\gamma_2\in\Gamma \\ k\in K}}\chi_{R_i}(k^{-1})Z_{\gamma_1,\gamma_2,\sigma_k},
\ee
where $Z_{\gamma_1,\gamma_2,\sigma_k}$ represents a correlation function taken on a torus with $\gamma_1$ and $\gamma_2$ TDLs wrapping the cycles and the TPO $\sigma_k$ inserted at their intersection, as shown in Figure~\ref{fig:mg}.  Analyzing these partition functions will be made easier due to the fact that many of the $Z_{\gamma_1,\gamma_2,\sigma_k}$ will vanish.  Consistency with the group law gives us the condition
\be
\label{mg_commutation}
\gamma_1\gamma_2=k\gamma_2\gamma_1.
\ee

Let us examine what this construction yields when $\Gamma=K$, i.e.~the entire zero-form symmetry is trivially-acting.  To begin, assume $K$ is abelian.  Then the condition (\ref{mg_commutation}) is only satisfied for $k=1$.  Since $G$ is trivial, there is no action of $G$ on $K$ and hence each irreducible representation is its own orbit, getting rid of the $i$ sum.  Each character appearing in (\ref{mg_pf}) is evaluated on the identity, giving 1.  Since all of $\Gamma$ acts trivially, the remaining $Z_{\gamma_1,\gamma_2,\sigma_1}$ are all equivalent to the parent theory partition function $Z_{1,1,\sigma_1}$ and we have
\be
\frac{1}{|\Gamma|^2}\sum_{\gamma_1,\gamma_2\in \Gamma}Z_{1,1,\sigma_{1}}=Z_{1,1,\sigma_1}.
\ee

Now assume $\Gamma=K$ is non-abelian -- for ease, we will begin by examining the universe corresponding to the identity representation.  There will now be pairs $\gamma_1,\gamma_2$ that do not commute (hence $Z_{\gamma_1,\gamma_2,\sigma_1}$ will vanish), but for each such pair there is a non-vanishing contribution to the sum from $Z_{\gamma_1,\gamma_2,\sigma_{[\gamma_1,\gamma_2]}}$, so we once again end up with a single copy of the parent theory partition function.

Moving to a generic irreducible representation $R$, we have
\be
\label{mg_nab_pf}
\frac{\text{dim}R}{|\Gamma|^2}\sum_{\gamma_1,\gamma_2\in\Gamma}\chi_{R}([\gamma_1,\gamma_2]^{-1})Z_{\gamma_1,\gamma_2,\sigma_{[\gamma_1,\gamma_2]}}.
\ee
Once again each factor of $Z$ will reduce to the parent theory partition function (since the group is entirely trivially-acting), so we simply need to tally up the coefficients.  This requires us to evaluate
\be
\label{mg_calc1}
\sum_{\gamma_1,\gamma_2\in\Gamma}\chi_R([\gamma_1,\gamma_2]^{-1}).
\ee
To do so, we use a classical result of Frobenius that the number $N(\gamma)$ of pairs $\gamma_1,\gamma_2$ satisfying $[\gamma_1,\gamma_2]=\gamma$ is given by
\be
N(\gamma)=|\Gamma|\sum_{R}\frac{\chi_R(\gamma)}{\chi_R(1)}=|\Gamma|\sum_R\frac{\chi_R(\gamma)}{\text{dim}R}.
\ee
This allows us to rewrite (\ref{mg_calc1}) as
\be
\sum_{\gamma\in\Gamma}N(\gamma)\chi_R(\gamma^{-1})=|\Gamma|\sum_{\substack{\gamma\in\Gamma \\ R'\in\text{Irr}\Gamma}}\frac{\chi_R(\gamma^{-1})\chi_{R'}(\gamma)}{\text{dim}R'}=\frac{|\Gamma|^2}{\text{dim}R},
\ee
where we have used orthonormality of characters to arrive at the final result.  This calculation shows that (\ref{mg_nab_pf}) reduces to a single copy of the parent theory partition function, independent of $R$.  Of course, this is exactly the result we would have expected: when we gauge a totally trivially-acting symmetry, each universe in the decomposition has the same local spectrum as the parent theory, so the result of projecting onto a single copy would be (a single copy of) the parent theory partition function.\\

In cases where $\Gamma\neq K$ (so there is an effective symmetry in play), we cannot produce a general answer as easily, so instead we will examine a particular example: the symmetric group $S_3$ with trivially-acting $\Z_3$ subgroup.  Again we begin by examining the projector associated to the identity representation.  Since the $\Z_3$ subgroup of $S_3$ is also its commutator subgroup, we can once again find exactly one $Z_{\gamma_1,\gamma_2,\sigma{[\gamma_1,\gamma_2]}}$ for each pair $\gamma_1,\gamma_2$.  The sum in (\ref{mg_pf}) will then have the maximum number of contributions (thirty six).  If we let $g$ be the generator of the effective $S_3/\Z_3\simeq\Z_2$ symmetry, (\ref{mg_pf}) in this case becomes
\be
\label{s3z3mg1}
\frac{1}{2}[Z_{1,1}+Z_{1,g}+Z_{g,1}+Z_{g,g}],
\ee
which is the partition function of a single copy of the $\Z_2$ orbifold of the corresponding theory with no trivially-acting symmetry.  Again this is in line with expectation, since the $S_3$ orbifold with trivially-acting $\Z_3$ decomposes into a copy of the parent theory plus a copy of its $\Z_2$ orbifold.

Of course, we should also be able to obtain the other universe present in the decomposition from mixed gauging.  The remaining two non-identity irreducible representations are exchanged by the action of $S_3/\Z_3\simeq\Z_2$ on $\Z_3$, so the $i$ sum in (\ref{mg_pf}) has two terms.  The relevant coefficients come from the projector which is built as
\be
\frac{1}{3}(\sigma_1+e^{2\pi i/3}\sigma_b+e^{4\pi i/3}\sigma_{b^2})+\frac{1}{3}(\sigma_1+e^{4\pi i/3}\sigma_b+e^{2\pi i/3}\sigma_{b^2})=\frac{1}{3}(2\sigma_1-\sigma_b-\sigma_{b^2}),
\ee
where $b$ generates the $\Z_3$ orbifold group.  Evaluating (\ref{mg_pf}) with these coefficients causes each contribution to the partition function with a $g$ to cancel out, and we are left with a single copy of the (unorbifolded) parent theory partition function $Z_{1,1}$, which is precisely the other universe in the decomposition.

In this context, the mixed anomalies of section~\ref{sec:mixedanom} are exactly what their name suggests: 
an obstruction to the mixed gauging of TDLs and TPOs.  
As the examples of the following section will make clear, the phases (\ref{mixedanomphases}) introduced by a mixed anomaly into the $[T/\Gamma_{[0]}]$ partition function obstruct its decomposition, which would prevent us from consistently being able to form projectors.

\section{Examples}
\label{sec:examples}

Using the framework developed in the previous section, we now examine TDLs and TPOs in examples of orbifolds by non-effectively-acting symmetries.  We will pay particular attention to the fate of TPOs under gauging to help us track the symmetries possessed by each theory.  We introduce the notation $T(\Gamma,K)$ to denote a theory $T$ with distinguished TDLs corresponding to total zero-form symmetry $\Gamma$ of which the normal subgroup $K \subset \Gamma$ acts trivially.  

The key idea in each example will be to track TPOs which are not associated to vacua.  For one simple example, consider decomposition in an orbifold $[X/\Gamma]$ where the abelian group $\Gamma$ is a central extension
\begin{equation}
1 \: \longrightarrow \: K \: \longrightarrow \: \Gamma \: \longrightarrow
\: G \: \longrightarrow \: 1
\end{equation}
by trivially-acting (and central) $K$.  In the absence of mixed anomalies or discrete torsion, the statement of decomposition~(\ref{eq:decomp:genl}) simplifies to the form
\begin{equation}
{\rm QFT}\left( [T/\Gamma] \right) \: = \:
{\rm QFT}\left( \coprod_{\rho \in \hat{K}} [X/G]_{\hat{\omega}(\rho)}
\right).
\end{equation}
Here, there are as many universes ($|K|$) as topological operators, so there are no `excess' TPOs, and so no additional trivially-acting symmetries are expected amongst the universes of the decomposition.

The story will become more interesting when the number of universes is smaller than the order of $K$.  This can happen, for instance, if there are phases in the partition function which cause sectors to cancel with each other upon decomposition.  These could be present in the $\Gamma$ orbifold from the beginning (discrete torsion in $\Gamma$), or could appear as part of the decomposition process (due to a mixed anomaly).  The other way to reduce the number of universes is for $\Gamma$ to be non-abelian.  Then a number of sectors $Z_{\gamma_1,\gamma_2}$ will simply be absent from the start, since there will be pairs $\gamma_1,\gamma_2$ that do not commute.  In each of these cases, there will be (linear combinations of) TPOs which do not form vacua, and instead remain bound to TDLs, providing some universes with trivially-acting symmetries.  Exploring how the TPOs distribute themselves among universes will be the focus of this section.

\subsection{$\Gamma=\Z_2\times\Z_2$ or $\Z_4, K=\Z_2$}

We begin by considering a theory $T(\Z_2,1)$, with an effectively-acting $\Z_2$ symmetry.  There exist two ways to supplement $T(\Z_2,1)$ with a trivially-acting $\Z_2$ symmetry, denoted $T(\Z_2\times\Z_2,\Z_2)$ and $T(\Z_4,\Z_2)$, both of which have the same local spectrum as $T$.  In $T(\Z_2\times\Z_2,\Z_2)$ the two symmetries combine trivially to give $\Gamma=\Z_2\times\Z_2$, while we take $T(\Z_4,\Z_2)$ to have non-trivial extension class, giving $\Gamma=\Z_4$.

As local theories, decomposition predicts that in both of these cases the $\Gamma$ orbifolds
are identical:
\begin{equation}
{\rm QFT}\left( [T / {\mathbb Z}_2 \times {\mathbb Z}_2] \right)
\: = \: {\rm QFT} \left( \coprod_2 [T/{\mathbb Z}_2] \right)
\: = \:
{\rm QFT}\left( [T/{\mathbb Z}_4] \right).
\end{equation}
as is easily checked by e.g.~partition functions.  However, one might also ask about extended objects in these theories.  Schematically, because there are no further TPOs beyond the vacua of the two universes, we do not expect either universe to have trivially-acting symmetries, hence
\begin{eqnarray}  \label{z2z2decomp}
\left[ T(\Z_2\times\Z_2,\Z_2)/(\Z_2\times\Z_2) \right] & = &
\left[ T(\Z_2,1)/\Z_2 \right] \, \coprod \, \left[  T(\Z_2,1)/\Z_2 \right],
\\
& = & \left[ T(\Z_4,\Z_2)/\Z_4 \right].
\end{eqnarray}

While the resulting theories are given by the same direct sum, there is additional information not captured in the notation above that distinguishes them.  We know to expect a $\Gamma$ orbifold to have a $\hat{\Gamma}$ quantum symmetry.  This means that (\ref{z2z2decomp}) should possess a $\Z_2\times\Z_2$ symmetry.  This is given by the product of the symmetry that acts as the effective $\Z_2$ on both copies with the $\Z_2$ exchange symmetry possessed by the two copies.  The story is similar for $[T(\Z_4,\Z_2)/\Z_4]$, except we expect a quantum $\Z_4$ symmetry.  In this case the two symmetries are combined non-trivially -- schematically, we have $(\text{exchange }\Z_2.\text{effective }\Z_2)\simeq\Z_4$.

In both cases above, the TPO that implements the trivially-acting $\Z_2$ symmetry in the ungauged theory becomes the vacuum of a second universe in the gauged theory.  Let us see how this changes if we take these theories to have non-trivial mixed anomaly.  From the analysis of sections \ref{sec:mixedanom} and \ref{sec:orbifolds}, we expect mixed anomalies for these theories to be classified by elements of $H^1(\Gamma,\hat{K})$ which are pulled back from $H^1(G,\hat{K})$, which in this case is simply $\text{Hom}(\Z_2,\Z_2)=\Z_2$.  Therefore there is one non-trivial choice of mixed anomaly.

Such mixed anomaly phases were denoted `quantum symmetry phases' in \cite{Robbins:2021ibx}, and from the decomposition statement in \cite[section 3]{Robbins:2021ibx}, we expect that for a $\Gamma = {\mathbb Z}_2 \times {\mathbb Z}_2$ orbifold, with $G = K = {\mathbb Z}_2$ and non-trivial $B \in H^1(G,\hat{K}) = {\mathbb Z}_2$ 
corresponding to an isomorphism $G \rightarrow K$, we have
\begin{equation}  \label{eq:z2z2-ma}
{\rm QFT}\left( [T / {\mathbb Z}_2 \times {\mathbb Z}_2] \right)
\: = \:
{\rm QFT}\left( \left[ \frac{ T \times \hat{ {\rm Coker}\, B } }{ {\rm Ker}\, 
B } \right] \right) \: = \: {\rm QFT}(T),
\end{equation}
as ${\rm Ker}\, B = 1 = {\rm Coker}\, B$.

Now, let us refine this statement to include TDLs and TPOs.  With this mixed anomaly,  our prescription (\ref{mixedanomphases}) for computing orbifold partition functions suggests that we reobtain the parent theory:
\be
\left[ T(\Z_2\times\Z_2,\Z_2)/(\Z_2\times\Z_2)|_{\text{M.A.}} \right]
\: = \:
T(\Z_2\times\Z_2,\Z_2)
\ee
(refining~(\ref{eq:z2z2-ma})) and
\be
\left[ T(\Z_4,\Z_2)/\Z_4|_{\text{M.A.}} \right]
\: = \:
T(\Z_4,\Z_2).
\ee

Note that in the $\Z_2\times\Z_2$ orbifold we could turn on discrete torsion, as well, and decomposition in orbifolds with discrete torsion was discussed in \cite{Robbins:2020msp}.  The phases that would result from this addition are in fact equivalent to those of the mixed anomaly, as discussed in \cite{Robbins:2021ibx}.  Using the results there, decomposition predicts
\begin{equation}
{\rm QFT} \left( [T / {\mathbb Z}_2 \times {\mathbb Z}_2] |_{\rm D.T.} \right)
\: = \:
{\rm QFT}(T),
\end{equation}
see e.g.~\cite[section 5.1]{Robbins:2020msp}.

If we refine the analysis to track trivially-acting symmetries, then from the same reasoning as above, we have
\be
\left[ T(\Z_2\times\Z_2,\Z_2)/(\Z_2\times\Z_2)|_{\text{D.T.}} \right]
\: = \:
T(\Z_2\times\Z_2,\Z_2).
\ee
In this case the mixed anomaly and discrete torsion have redundant effects.  In the $\Z_4$ orbifold, of course, there is no choice of discrete torsion, as $H^2(\Z_4,U(1))$ is trivial.  The phases from the mixed anomaly are something wholly dependent on the presence of a trivially-acting subgroup.

\subsection{$\Gamma=S_3, K=\Z_3$}
\label{sec:s3z3}

Next we take on the simplest non-abelian example, a theory with TDLs that fuse as the symmetric group $S_3$ with a trivially-acting, non-central $\Z_3$ subgroup.  We will present $S_3$ as $\braket{a,b|a^2=b^3=1,abab=1}$.  The ungauged theory has three TPOs, which we can label $\sigma_1, \sigma_b, \sigma_{b^2}$, and using the fact that $G = {\mathbb Z}_2$ exchanges two irreducible representations of $K = {\mathbb Z}_3$, leaving the third invariant, from~(\ref{eq:decomp:genl}) we see that decomposition predicts 
\begin{equation}  
\label{eq:decomp:s3-z3}
{\rm QFT}\left( [T/S_3] \right) \: = \:  {\rm QFT}\left( 
[T/{\mathbb Z}_2] \, \coprod \, T \right).
\end{equation}

Now, let us consider extended objects in this decomposition.  The algebra object for gauging is simply given by the direct sum of the group elements $A=1+a+b+ab+b^2+ab^2$.  We can quickly determine how many bulk TPOs exist in the gauged theory by acting on each ungauged TPO with $A$:
\begin{align}
\frac{1}{6}A\cdot\sigma_1&=\sigma_1\\
\frac{1}{6}A\cdot\sigma_b&=\frac{1}{2}(\sigma_b+\sigma_{b^2})\\
\frac{1}{6}A\cdot\sigma_{b^2}&=\frac{1}{2}(\sigma_b+\sigma_{b^2}).
\end{align}
Due to the non-trivial action of $\Z_2$ on $\Z_3$, we find that there are only two distinct bulk TPOs, consistent with the decomposition into two universes described above.

Next, we compute projectors.  From~(\ref{eq:proj}), corresponding to the three irreducible representations of $K = {\mathbb Z}_3$ we have the three constituents
\begin{eqnarray}
\Pi_0 & = & \frac{1}{3} \left( \sigma_1 + \sigma_b + \sigma_{b^2} \right),
\\
\Pi_1 & = & \frac{1}{3} \left( \sigma_1 + \xi \sigma_b + \xi^2 \sigma_{b^2}
\right),
\\
\Pi_2 & = & \frac{1}{3} \left( \sigma_1 + \xi^2 \sigma_b + \xi^2 \sigma_{b^2}
\right).
\end{eqnarray}
The universe $[T/{\mathbb Z}_2]$ corresponds to the trivial representation of ${\mathbb Z}_3$, and the universe $T$ corresponds to each of the two non-trivial representations.  If we denote $T$ by $L$ and $[T/{\mathbb Z}_2]$ by $R$, then we have the projectors
\begin{eqnarray}
\Pi_L &  = & \frac{1}{3}(2\sigma_1-\sigma_b-\sigma_{b^2})
\: = \: \Pi_1 + \Pi_2,\\
\Pi_R &= & \frac{1}{3}(\sigma_1+\sigma_b+\sigma_{b^2})
\: = \: \Pi_0.
\end{eqnarray}
Next, we consider the fate of the third TPO under gauging.  We can see the answer by defining a quantity
\be
g_L \: = \: \frac{i}{\sqrt{3}}(\sigma_b-\sigma_{b^2}) \: = \:
\Pi_1 - \Pi_2.
\ee
Note that we have $g_L\cdot \Pi_R=0$, meaning that $g_L$ is an operator localized to the $L$ universe.  Also, it satisfies the relations $g_L\cdot \Pi_L = \Pi_L\cdot g_L=g_L$ and $g_L\cdot g_L=\Pi_L$, i.e.~it obeys a $\Z_2$ fusion law with that universe's vacuum.  Thus, the third TPO is bound to TDLs in the $L$ universe, and implements a trivially-acting $\Z_2$ symmetry there.  In our schematic notation, the full decomposition including TDLs and TPOs is
\be
\left[ T(S_3,\Z_3)/S_3 \right]
 \: = \:
 T(\Z_2\times\Z_2,\Z_2) \, \coprod \, \left[ T(\Z_2,1)/\Z_2 \right],
\ee
which refines~(\ref{eq:decomp:s3-z3}).

\subsection{$\Gamma=Q_8, K=\Z_4$}
\label{sec:q8ex}

In this example we once again have a non-central subgroup, $\Z_4$ in the group $Q_8$ of unit quaternions.  We present $Q_8$ in the usual manner as 
\begin{equation}
\{ i,j,k \,| \, i^2 = j^2 = k^2 = ijk = -1 \}.  
\end{equation}
Taking the trivially-acting $\Z_4$ to be generated by $k$, there are four TPOs in the ungauged theory: $\sigma_1,\sigma_k,\sigma_{-1}$ and $\sigma_{-k}$.  The decomposition of the $Q_8$ orbifold was discussed in
\cite[section 5.4]{Hellerman:2006zs}, where it was argued that
\begin{equation}  
\label{eq:decomp:q8-z4}
{\rm QFT}\left( [T/Q_8] \right) \: = \:
{\rm QFT}\left( [T/{\mathbb Z}_2] \coprod [T/{\mathbb Z}_2] \coprod T \right).
\end{equation}
This can also be seen from the formula~(\ref{eq:decomp:genl}), using the fact that of the four irreducible representations of the trivially-acting ${\mathbb Z}_4$, two are invariant under $G = {\mathbb Z}_2$ (corresponding to the two copies of $[T/{\mathbb Z}_2]$) and the other two are exchanged (corresponding to the one copy
of $T$).  In this section we will analyze the decomposition of the corresponding TDLs and TPOs, refining the decomposition statement above.

First, consider the algebra object $A$ for the gauging above, which is the sum of the eight group elements.  The action on the ungauged TPOs is readily verified to be
\begin{align}
\frac{1}{8}A\cdot \sigma_1&=\sigma_1, \\
\frac{1}{8}A\cdot \sigma_{-1}&=\sigma_{-1}, \\
\frac{1}{8}A\cdot \sigma_k&=\frac{1}{2}(\sigma_k+\sigma_{-k}), \\
\frac{1}{8}A\cdot \sigma_{-k}&=\frac{1}{2}(\sigma_k+\sigma_{-k}),
\end{align}
which signals that we should expect three universes in the decomposition.  Their vacua correspond to the projection operators from~(\ref{eq:proj}) (see also \cite[section 5.4.2]{Hellerman:2006zs},
\cite[section 4.2]{Sharpe:2021srf})
\begin{align}
\Pi_a&=\frac{1}{2}(\sigma_1-\sigma_{-1}), \\
\Pi_b&=\frac{1}{4}(\sigma_1+\sigma_{-1}+\sigma_k+\sigma_{-k}), \\
\Pi_c&=\frac{1}{4}(\sigma_1+\sigma_{-1}-\sigma_k-\sigma_{-k}).
\end{align}

Based on the previous examples we would expect the fourth TPO to have become localized to one of the three universes.  Indeed, the quantity
\be
g_a=\frac{i}{2}(\sigma_k-\sigma_{-k})
\ee
lives in the $a$ universe and generates a trivially-acting $\Z_2$ symmetry there.  The full decomposition at the level of TDLs and TPOs then takes the form
\be
\left[ T(Q_8,\Z_4)/Q_8 \right]
\: = \:
T(\Z_2\times\Z_2,\Z_2) \, \coprod \,
\left[ T(\Z_2,1)/\Z_2 \right] \, \coprod \,
\left[ T(\Z_2,1)/\Z_2 \right],
\ee
refining the decomposition described in \cite[section 5.4]{Hellerman:2006zs} and above in~(\ref{eq:decomp:q8-z4}) by telling us that the universe corresponding to the unorbifolded theory carries an extra trivially-acting symmetry.

We can also take into account a mixed anomaly in this example.  The zero-form symmetry fits into the short exact sequence 
\begin{equation}
1 \: \longrightarrow \: {\mathbb Z}_4 \: \longrightarrow \: Q_8
\: \longrightarrow \: {\mathbb Z}_2 \: \longrightarrow \: 1,
\end{equation}
and $H^1(\Z_2,\hat{\Z}_4) = \Z_2$, with non-trivial action on the coefficients.  Decomposition in this example with non-trivial mixed anomaly was discussed in \cite[section 5.1.3]{Robbins:2021lry}, where it was argued that 
\begin{equation}  
\label{eq:decomp:q8-ma}
{\rm QFT}\left( [T/Q_8] \right) \: = \:
{\rm QFT}\left( T \, \coprod \, T \right).
\end{equation}
We shall next track TDLs and TPOs through this decomposition.

Pulling back the non-trivial element of $H^1(\Z_2,\hat{\Z}_4)$ to an element of $H^1(Q_8,\hat{\Z}_4)$ and using that to modify the action of the TDLs on the TPOs, we find the bulk gauged TPOs to be
\begin{align}
\frac{1}{8}A\cdot \sigma_1&=\sigma_1, \\
\frac{1}{8}A\cdot \sigma_{-1}&=0, \\
\frac{1}{8}A\cdot \sigma_{k}&=\frac{1}{2}(\sigma_k+i\sigma_{-k}), \\
\frac{1}{8}A\cdot \sigma_{-k}&=\frac{1}{2i}(\sigma_k+i\sigma_{-k}).
\end{align}

Since we now only have two independent combinations of TPOs, we expect a decomposition into two universes.  The two vacua are given by
\be
\Pi_{\pm}=\frac{1}{2}\left[\sigma_1\pm\left(\frac{1-i}{2}\sigma_k+\frac{1+i}{2}\sigma_{-k}\right)\right].
\ee
Each universe carries a single TPO beyond the vacuum, hence a distinguished TDL describing a trivially-acting $\Z_2$ symmetry controlled by
\be
g_{\pm}\: = \: \frac{1}{2\sqrt{2i}}[(1+i)\sigma_{-1}\pm i\sigma_k\pm \sigma_{-k}].
\ee
The full decomposition, then, takes the form 
\be
\left[ T(Q_8,\Z_4)/Q_8|_{\text{M.A.}} \right]
\: = \:
T(\Delta,\Z_2) \, \coprod \, T(\Delta,\Z_2),
\ee
where $\Delta$ is either ${\mathbb Z}_2 \times {\mathbb Z}_2$ or ${\mathbb Z}_4$.  This refines the decomposition result from \cite[section 5.1.3]{Robbins:2021lry} and above in~(\ref{eq:decomp:q8-ma}).

\subsection{$\Gamma=K=S_3$}
\label{sect:ex:s3}

In this example we take a theory with TDLs describing a completely trivially-acting $S_3$ symmetry.  Once again there are TPOs $\sigma_i$ in the ungauged theory for each trivially-acting group element.  The qualitative difference here is that since $K$ is non-abelian, the trivial symmetries in the post-decomposition universes can be non-abelian as well.  
As also discussed in section~\ref{sec:decomp}, after gauging the trivially-acting $S_3$, one gets three copies of the original theory:
\begin{equation} 
\label{eq:decomp:s3-s3}
{\rm QFT}\left( [T/S_3] \right)
\: = \:
{\rm QFT}\left( T \, \coprod \, T \, \coprod T \right),
\end{equation}
corresponding to the fact that there are three irreducible representations of $S_3$.

\begin{table}[ht]
\begin{center}
\begin{tabular}{crrr}
Representation & $\{1\}$ & $\{b, b^2\}$ & $\{a, ab, ab^2\}$ \\ \hline
$1$ & $1$ & $1$ & $1$ \\
$X$ & $1$ & $1$ & $-1$ \\
$Y$ & $2$ & $-1$ & $0$
\end{tabular}
\caption{\label{table:s3:chars}
Character table for $S_3$.
}
\end{center}
\end{table}

Labeling the projectors into these universes as $\Pi_1$, $\Pi_X$, $\Pi_Y$, from the general formula~(\ref{eq:proj}) and the character table~\ref{table:s3:chars} we have the projectors
\begin{eqnarray}
\Pi_1 & = & \frac{1}{6} \left( \sigma_1 \: + \: \sigma_{b} \: + \:
\sigma_{b^2} \: + \:
\sigma_{a} \: + \: \sigma_{ab} \: + \: \sigma_{ab^2} \right),
\\
\Pi_X & = & \frac{1}{6} \left( \sigma_1 \: + \: \sigma_{b} \: + \:
\sigma_{b^2} \: - \:
\sigma_{a} \: - \: \sigma_{ab} \: - \: \sigma_{ab^2} \right),
\\
\Pi_Y & = & \frac{2}{6} \left( 2\sigma_1 \: - \: \sigma_{b} \: - \:
\sigma_{b^2} \right).
\end{eqnarray}
In passing, we can use the explicit expressions above and a product of the form
\begin{equation}
R \cdot \sigma \: = \: \chi_R(\sigma) \sigma,
\end{equation}
for any representation $R$ and corresponding character $\chi_R$, corresponding to the quantum symmetry, to justify the products given earler in equations~(\ref{reps3-1})-(\ref{reps3-4}).

The three remaining linearly independent combinations form TPOs local to the $Y$ universe, which take the form
\begin{eqnarray}
i_Y & = & \frac{1}{\sqrt{3}}(\sigma_b-\sigma_{b^2}),
\\
j_Y & = & \frac{i}{\sqrt{3}}(\sigma_{ab}-\sigma_{ab^2}),
\\
k_Y & = & \frac{i}{3}[2\sigma_a-(\sigma_{ab}+\sigma_{ab^2})].
\end{eqnarray}
These can be checked to satisfy 
\begin{equation}
i_Y^2 \: = \: j_Y^2 \: = \: k_Y^2 \: = \: i_Yj_Yk_Y \: = \: -\Pi_Y, 
\end{equation}
group relations which closely resemble those of $Q_8$.  However, as $-\Pi_Y$ is not a distinct operator from $\Pi_Y$, 
this should be understood as the non-trivial projective representation of $\Z_2\times\Z_2$.  The decomposition including explicitly the distinguished TDLs and TPOs is then
\be
\label{s3decomp}
\left[ T(S_3,S_3)/S_3 \right] \: = \:
T(1,1) \, \coprod \,
 T(1,1) \, \coprod \,
 T(\Z_2\times\Z_2,\Z_2\times\Z_2)_{\text{proj.}},
\ee
which refines the decomposition example in section~\ref{sec:decomp} and above in~(\ref{eq:decomp:s3-s3}).

This example is an ideal illustration of the additional information gained by examining decomposition with TPOs in mind: the traditional partition function analysis of section~\ref{sec:decomp} told us that gauging a trivially-acting $S_3$ symmetry gives us three copies of the parent theory.  While it is true that all three theories in the direct sum (\ref{s3decomp}) have identical local spectra, we see that they do in fact differ in their extended spectra.  This also helps make sense of the asymmetric manner (\ref{reps3-1})-(\ref{reps3-4}) in which the fusion category $\text{Rep}(S_3)$ acts on the gauged theory.\\

This has also been our first example of TPOs which fuse projectively, a phenomenon which is quite readily understandable in the context of discrete torsion.  Recall that for an effective group-like symmetry $K$, networks of TDLs will decompose into three-way junctions, where e.g.~a $k_1k_2$ line splits into $k_1$ and $k_2$.  The Hilbert space of topological operators at this junction is isomorphic to $\C$.  Whatever phase sits at this junction, however, is not unique -- we are free to redefine the contribution of a line labeled by $k$ by a cochain $\lambda:K\to U(1)$, which we will take to obey $\lambda^{-1}(k)=\lambda(k^{-1})$.  Doing so changes the junction phase by $\lambda(k_1k_2)\lambda^{-1}(k_1)\lambda^{-1}(k_2)$, which we recognize as a coboundary shift.  Thus the phases at such junctions are meaningful up to their class in $H^2(K,U(1))$, and this choice provides the discrete torsion when we gauge $K$.

\begin{figure}
	\begin{subfigure}{0.5\textwidth}
		\centering
		\begin{tikzpicture}
		\draw[thick,->] (0,0) -- (0,0.5);
		\draw[thick] (0,0.5) -- (0,1);
		\node at (0.6,0.2) {$k_1k_2$};
		\filldraw[black] (0,1) circle (2pt);
		\node at (1,0.9) {$\sigma_{(k_1k_2)^{-1}}$};
		\draw[thick,->] (1,2) -- (1.5,2.5);
		\draw[thick] (1.5,2.5) -- (2,3);
		\filldraw[black] (1,2) circle (2pt);
		\node at (1,1.5) {$\sigma_{k_2}$};
		\node at (1.25,2.75) {$k_2$};
		\draw[thick,->] (-1,2) -- (-1.5,2.5);
		\draw[thick] (-1.5,2.5) -- (-2,3);
		\filldraw[black] (-1,2) circle (2pt);
		\node at (-1,1.5) {$\sigma_{k_1}$};
		\node at (-1.25,2.75) {$k_1$};
		\end{tikzpicture}
		\caption{}
		\label{dtproj1}
	\end{subfigure}
	\begin{subfigure}{0.5\textwidth}
		\centering
		\begin{tikzpicture}
		\draw[thick,->] (0,0) -- (0,0.5);
		\draw[thick] (0,0.5) -- (0,1);
		\node at (0.6,0.2) {$k_1k_2$};
		\filldraw[black] (0,1) circle (2pt);
		\node at (0.9,0.9) {$\sigma_{(k_1k_2)^{-1}}$};
		\draw[thick,->] (0,2) -- (0.5,2.5);
		\draw[thick] (0.5,2.5) -- (1,3);
		\filldraw[black] (0,2) circle (2pt);
		\node at (1.6,1.8) {$\sigma_{k_1k_2}\times\omega(k_1,k_2)$};
		\node at (1,2.5) {$k_2$};
		\draw[thick,->] (0,2) -- (-0.5,2.5);
		\draw[thick] (-0.5,2.5) -- (-1,3);
		\node at (-1,2.5) {$k_1$};
		\end{tikzpicture}
		\caption{}
		\label{dtproj2}
	\end{subfigure}
	\begin{subfigure}{\textwidth}
		\centering
		\begin{tikzpicture}
		\draw[thick,->] (0,0) -- (0,0.5);
		\draw[thick] (0,0.5) -- (0,1);
		\node at (0.6,0.2) {$k_1k_2$};
		\filldraw[black] (0,1) circle (2pt);
		\node at (1.5,0.9) {$\sigma_1\times\omega(k_1,k_2)$};
		\draw[thick,->] (0,1) -- (0.5,1.5);
		\draw[thick] (0.5,1.5) -- (1,2);
		\node at (1,1.5) {$k_2$};
		\draw[thick,->] (0,1) -- (-0.5,1.5);
		\draw[thick] (-0.5,1.5) -- (-1,2);
		\node at (-1,1.5) {$k_1$};
		\end{tikzpicture}
		\caption{}
		\label{dtproj3}
	\end{subfigure}
	\caption{We build a three-way junction out of TPOs that fuse projectively.}
	\label{fig:dtproj}
\end{figure}
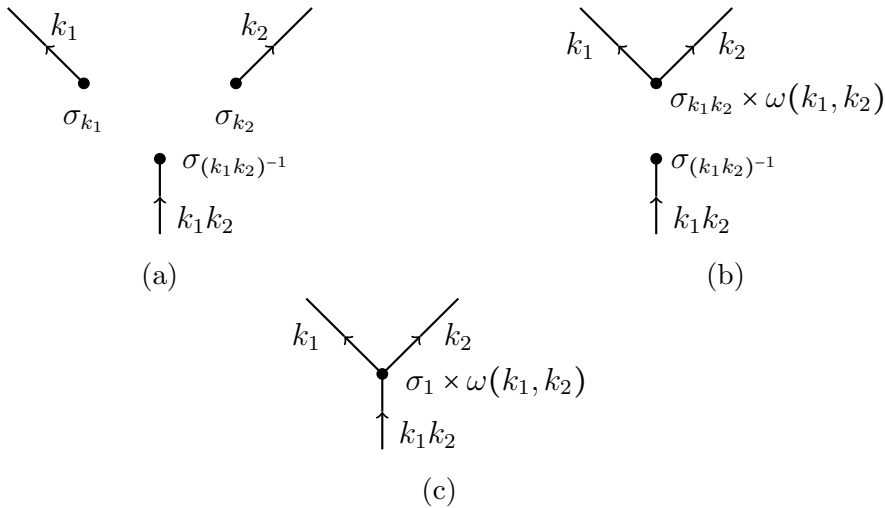

Now consider the case where $K$ is trivially-acting.  We could now build the three-way junction described above out of lines that end on TPOs, as illustrated in Figure~\ref{fig:dtproj}.  The initial setup is shown in Figure~\ref{dtproj1}.  In Figure~\ref{dtproj2} we fuse the $\sigma_{k_1}$ and $\sigma_{k_2}$ TPOs, allowing the possibility that they are in a projective representation of $K$, such that the result $\sigma_{k_1k_2}$ is multiplied by some $\omega(k_1,k_2)\in H^2(K,U(1))$.  In Figure~\ref{dtproj3} we fuse $\sigma_{k_1k_2}$ and $\sigma_{(k_1k_2)^{-1}}$ to the identity TPO.  In principle this could produce another phase $\omega(k_1k_2,(k_1k_2)^{-1})$, but our earlier insistence to work with cochains satisfying $\lambda^{-1}(k)=\lambda(k^{-1})$ is equivalent to the condition that $\omega(k,k^{-1})=1$ (and there is always such a choice of representative).  Therefore we end up with the situation described above, where a $k_1k_2$ line splits into $k_1$ and $k_2$ and the junction is multiplied by a phase given by a class in $H^2(K,U(1))$.  This demonstrates that projective fusion of TPOs corresponding to trivially-acting symmetries is equivalent to turning on discrete torsion in that symmetry.  Note the similarity here to the correspondence between symmetry-protected topological (SPT) phases and anomalies of theories on the boundary \cite{Wen_2013}.

\subsection{$\Gamma=K=\text{Rep}(S_3)$}

In our final example, we look at a theory with a trivially-acting fusion categorical (as opposed to group-like) symmetry.  We can construct such a theory by beginning with the direct sum of $n!$ copies of a theory.  These copies should have a zero-form $S_n$ exchange symmetry.  Gauging this should bring us to a 
single\footnote{
We identify each copy with an element of $S_n$, and then let $S_n$ act
by left multiplication (i.e.~the copies transform in the regular
representation).  The action is both free and transitive, 
so the quotient is a single
copy.  Since the copies are disjoint, there can be no twisted sectors,
so the action of the quantum symmetry is trivial.
} copy of the theory, which we expect on general grounds to have a quantum symmetry given by the fusion category Rep$(S_n)$ \cite{Bhardwaj:2017xup}.  As we have seen in previous examples, as the quantum dual to an exchange symmetry, this Rep$(S_n)$ should act `trivially' on the theory (and we will shortly see how one adapts the notion of a trivial action to a non-group-like scenario).

The simplest non-trivial example of such a construction occurs for $n=3$, where our theory has a trivially-acting Rep$(S_3)$ symmetry.  This fusion category was used as a running example throughout section~\ref{sec:background}; its fusion products were given in Figure~\ref{fig:reps3}.  Such a theory has three simple TDLs labeled by $1$, $X$ and $Y$.  $1$ and $X$, which form a $\Z_2$ subsymmetry of Rep$(S_3)$, have the familiar trivial action on local operators $\mcO$:
\begin{align}
1\cdot\mcO&=\mcO, \\
X\cdot\mcO&=\mcO.
\end{align}
The remaining simple line $Y$, which is of dimension 2, acts on operators as
\be
Y\cdot\mcO=2\mcO.
\ee
The easiest way to see that this must be the case is to note that this `trivial' action should still respect the fusion rules, which in particular means that $(Y\otimes Y)\cdot\mcO=(1+X+Y)\cdot\mcO$.  In general, taking a line of dimension $d$ to act as multiplication by $d$ will respect the fusion rules 
(and this agrees with the notion that in the group-like case we can take all trivially-acting TDLs to act as multiplication by 1).  Additionally, taking $\mcO$ trivial tells us that an empty loop labeled by a line $A$ contributes a factor of its dimension $d_A$ to a correlation function, which is also a standard result \cite{Fuchs:2002cm}.

As in group-like examples, the triviality of this symmetry should mean that there exist TPOs which live on the TDLs in our theory.  There should be TPOs $\sigma_1$ and $\sigma_X$ living at the end of identity and $X$ lines.  Since $Y$ is weight two, however, we should have $\text{dim}(\text{Hom}(Y,1))=2$, giving us two TPOs $\sigma_{Y_1}$ and $\sigma_{Y_2}$.

Gauging the Rep$(S_3)$ symmetry will entail selecting the algebra object $A=1+X+2Y$ (which corresponds to the regular representation of $S_3$).  As before, the vacua of the various universes in the gauged theory should be expressible as linear combinations of the TPOs from the ungauged theory.  This example will prove no exception.  Note that because two copies of $Y$ appear in the algebra object, we will have four $Y$ TPOs.  Denoting the $Y$ representations as $Y^1$ and $Y^2$, then, the six TPOs present in the ungauged theory are $\sigma_1, \sigma_X, \sigma_{Y^1_1}, \sigma_{Y^1_2}, \sigma_{Y^2_1}, \sigma_{Y^2_2}$.  We can refer to \cite[section 4.6.1]{bhardwaj2023gapped} for their fusion and the resulting vacuum structure, which (given an implicit choice of basis which mixes the four operators coming from the $Y$ lines) are
\be
\label{rs3_tpo_fusion}
\begin{tabular}{| c | c | c | c | c | c | c |}
\hline
$\otimes$ & $\sigma_1$ & $\sigma_X$ & $\sigma_{Y^1_1}$ & $\sigma_{Y^1_2}$ & $\sigma_{Y^2_1}$ & $\sigma_{Y^2_2}$ \\
\hline
$\sigma_1$ & $\sigma_1$ & $\sigma_X$ & $\sigma_{Y^1_1}$ & $\sigma_{Y^1_2}$ & $\sigma_{Y^2_1}$ & $\sigma_{Y^2_2}$ \\
\hline
$\sigma_X$ & $\sigma_X$ & $\sigma_1$ & $\sigma_{Y^1_1}$ & -$\sigma_{Y^1_2}$ & -$\sigma_{Y^2_1}$ & $\sigma_{Y^2_2}$ \\
\hline
$\sigma_{Y^1_1}$ & $\sigma_{Y^1_1}$ & $\sigma_{Y^1_1}$ & $\sigma_{Y^2_2}$ & 0 & 0 & $\frac{1}{2}(\sigma_1+\sigma_X)$\\
\hline
$\sigma_{Y^1_2}$ & $\sigma_{Y^1_2}$ & -$\sigma_{Y^1_2}$ & 0 & $\sigma_{Y^2_1}$ & $\frac{1}{2}(\sigma_1-\sigma_X)$ & 0 \\
\hline
$\sigma_{Y^2_1}$ & $\sigma_{Y^2_1}$ & -$\sigma_{Y^2_1}$ & 0 & $\frac{1}{2}(\sigma_1-\sigma_X)$ & $\sigma_{Y^1_2}$ & 0 \\
\hline
$\sigma_{Y^2_2}$ & $\sigma_{Y^2_2}$ & $\sigma_{Y^2_2}$ & $\frac{1}{2}(\sigma_1+\sigma_X)$ & 0 & 0 & $\sigma_{Y^1_1}$\\\hline
\end{tabular}
\ee
and
\begin{align}
\Pi_1 &= \frac{1}{6}(\sigma_1+\sigma_X+2\sigma_{Y^1_1}+2\sigma_{Y^2_2})\\
\Pi_2 &= \frac{1}{6}(\sigma_1+\sigma_X+2\omega\sigma_{Y^1_1}+2\omega^2\sigma_{Y^2_2})\\
\Pi_3 &= \frac{1}{6}(\sigma_1+\sigma_X+2\omega^2\sigma_{Y^1_1}+2\omega\sigma_{Y^2_2})\\
\Pi_4 &= \frac{1}{6}(\sigma_1-\sigma_X+2\sigma_{Y^1_2}+2\sigma_{Y^2_1})\\
\Pi_5 &= \frac{1}{6}(\sigma_1-\sigma_X+2\omega\sigma_{Y^1_2}+2\omega^2\sigma_{Y^2_1})\\
\Pi_6 &= \frac{1}{6}(\sigma_1-\sigma_X+2\omega^2\sigma_{Y^1_2}+2\omega\sigma_{Y^2_1})
\end{align}
where $\omega=\exp{2\pi i/3}$.  One can check that under the fusion (\ref{rs3_tpo_fusion}), the expressions above form orthonormal projectors.  Thus, under gauging the regular representation in Rep$(S_3)$, we recover the six universes from which we built the theory with trivially-acting Rep$(S_3)$ symmetry.
\section{Conclusion}

While two-dimensional theories proved convenient for the calculations in this paper, many of the statements regarding topological operators should apply to general dimension.  For instance, one can engineer decomposition in three-dimensional theories by gauging trivially-acting one-form symmetries \cite{Pantev:2022kpl}, and in general dimension by gauging a trivially-acting $(d-2)$-form symmetry.  Of course, the explanation in terms of topological operators is much the same in each of these cases: $(d-2)$-form symmetries are controlled by topological operators of codimension $d-1$, i.e.~TDLs.  The trivial action of these TDLs is controlled by a set of operators of codimension $d$ (TPOs) which live bound to those lines.  Gauging the $(d-2)$-form symmetry (potentially) ``frees'' the TPOs, which act as vacua of disjoint universes.  In dimensions above two, we have access to a greater variety of ingredients -- there will be various topological operators of dimension greater than one, and it would be interesting to extend the analysis of this paper to such cases, where there is potential for more varied mixed gaugings, mixed anomalies, etc.

Even in two dimensions, there remains work to be done.  The analysis of the examples of section~\ref{sec:examples} provides a refinement of the usual decomposition story, in which we track how the TPOs from the ungauged theory divide themselves among the universes.  This tells us not only how many universes we have, but provides information about the symmetries that are localized to each universe.  Does this provide a full characterization of the symmetries of each universe, or are there further refinements available?  Also, are we able to create heuristics to predict how these TPOs should split without doing the full computation, as we are for e.g.~which symmetries are gauged in each universe \cite{Hellerman:2006zs,Robbins:2020msp}?  Finally, we could consider how the story changes when our zero-form symmetries have gauge anomalies.  These are points we plan to pursue in future work.

\section*{Acknowledgements}

We would like to thank D.~Ben-Zvi, T.~Pantev, and Y.~Tachikawa 
for useful discussions.
D.R.~was partially supported by
NSF grant PHY-1820867.
E.S.~was partially supported by NSF grant
PHY-2014086.

\appendix

\section{Consistency Conditions of $B$ for Bound TPOs}
\label{sec:boundb}

\begin{figure}
	\begin{subfigure}{0.5\textwidth}
		\centering
		\begin{tikzpicture}
		\draw[very thick,->] (2,0) -- (1,0);
		\draw[very thick] (1,0) -- (0,0);
		\node at (1,0.5) {$\gamma k\gamma^{-1}$};
		\draw[thick,->] (3,0) [partial ellipse=-90:90:1cm and 1cm];
		\node at (3,1.3) {$\gamma k$};
		\draw[thick,->] (3,0) [partial ellipse=90:270:1cm and 1cm];
		\node at (3,-1.3) {$\gamma$};
		\filldraw[black] (4,0) circle (2pt);
		\node at (4.4,0) {$\sigma_k$};
		\end{tikzpicture}
	\caption{}
	\label{boundactiona}
	\end{subfigure}
	\begin{subfigure}{0.5\textwidth}
		\centering
		\begin{tikzpicture}
		\draw[very thick,->] (2,0) -- (1,0);
		\draw[very thick] (1,0) -- (0,0);
		\node at (1,0.5) {$\gamma k\gamma^{-1}$};
		\filldraw[black] (2,0) circle (2pt);
		\node at (2.75,0) {$\sigma_{\gamma k\gamma^{-1}}$};
		\end{tikzpicture}
		\vspace{1cm}
	\caption{}
	\label{boundactionb}
	\end{subfigure}
	\label{fig:boundaction}
	\caption{The action of $\gamma$ on $\sigma_k$ for a bound TPO.}
\end{figure}
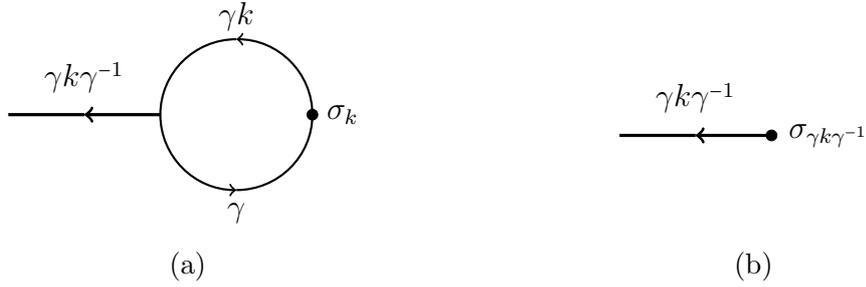

We repeat the analysis of section~\ref{sec:mixedanom} in the case that the TPOs are bound to lines.  In this case, the mixed anomaly phase $B(\gamma,\gamma k\gamma^{-1})$ would arise when moving from the configuration shown in Figure~\ref{boundactiona} to that of Figure~\ref{boundactionb}.  For readability of the figures, it will be convenient in this section to use a convention where a TPO labeled by $k$ multiplies the label of an incoming TDL from the right rather than the left.

\begin{figure}
	\begin{subfigure}{0.5\textwidth}
		\centering
		\begin{tikzpicture}
		\draw[very thick,->] (3,0) -- (1.5,0);
		\draw[very thick] (1.5,0) -- (0,0);
		\node at (1.5,0.5) {$\gamma k_1k_2 \gamma^{-1}$};
		\draw[thick,->] (4,0) [partial ellipse=0:90:1cm and 1cm];
		\node at (4,-1.3) {$\gamma$};
		\draw[thick,->] (4,0) [partial ellipse=90:270:1cm and 1cm];
		\node at (4,1.3) {$\gamma k_1k_2$};
		\filldraw[black] (4.7,0.7) circle (2pt);
		\node at (5.25,0.7) {$\sigma_{k_2}$};
		\draw[thick,->] (4,0) [partial ellipse=-90:0:1cm and 1cm];
		\node at (5.5,0) {$\gamma k_1$};
		\filldraw[black] (4.7,-0.7) circle (2pt);
		\node at (5.25,-0.7) {$\sigma_{k_1}$};
		\end{tikzpicture}
	\caption{}
	\label{boundkhoma}
	\end{subfigure}
	\begin{subfigure}{0.5\textwidth}
		\centering
		\begin{tikzpicture}
		\draw[very thick,->] (3,0) -- (1.5,0);
		\draw[very thick] (1.5,0) -- (0,0);
		\node at (1.5,0.5) {$\gamma k_1k_2\gamma^{-1}$};
		\draw[thick,->] (4,0) [partial ellipse=-90:90:1cm and 1cm];
		\node at (4,-1.3) {$\gamma$};
		\draw[thick,->] (4,0) [partial ellipse=90:270:1cm and 1cm];
		\node at (4,1.3) {$\gamma k_1k_2$};
		\filldraw[black] (5,0) circle (2pt);
		\node at (5.75,0) {$\sigma_{k_1k_2}$};
		\end{tikzpicture}
	\caption{}
	\label{boundkhomb}
	\end{subfigure}
	\begin{subfigure}{\textwidth}
		\centering
		\begin{tikzpicture}
		\draw[very thick,->] (3,0) -- (1.5,0);
		\draw[very thick] (1.5,0) -- (0,0);
		\node at (1.5,0.5) {$\gamma k_1 k_2\gamma^{-1}$};
		\draw[thick,->] (4,0) [partial ellipse=-90:90:1cm and 1cm];
		\node at (4,-1.3) {$\gamma$};
		\draw[thick,->] (4,0) [partial ellipse=90:270:1cm and 1cm];
		\node at (4,1.3) {$ \gamma k_1k_2$};
		\filldraw[black] (5,0) circle (2pt);
		\node at (5.3,-0.25) {$\sigma_{k_2}$};
		\draw[very thick,->] (8,0) -- (6.5,0);
		\draw[very thick] (6.5,0) -- (5,0);
		\node at (6.5,0.5) {$\gamma k_1\gamma^{-1}$};
		\draw[thick,->] (9,0) [partial ellipse=-90:90:1cm and 1cm];
		\node at (9,1.3) {$\gamma k_1$};
		\draw[thick,->] (9,0) [partial ellipse=90:270:1cm and 1cm];
		\node at (9,-1.3) {$\gamma$};
		\filldraw[black] (10,0) circle (2pt);
		\node at (10.5,0) {$\sigma_{k_1}$};
		\end{tikzpicture}
	\caption{}
	\label{boundkhomc}
	\end{subfigure}
	\begin{subfigure}{0.5\textwidth}
		\centering
		\begin{tikzpicture}
		\draw[very thick,->] (2,0) -- (1,0);
		\draw[very thick] (1,0) -- (0,0);
		\node at (1,0.5) {$\gamma k_1 k_2\gamma^{-1}$};
		\filldraw[black] (2,0) circle (2pt);
		\node at (2,-0.5) {$\sigma_{\gamma k_2\gamma^{-1}}$};
		\draw[very thick,->] (4,0) -- (3,0);
		\draw[very thick] (3,0) -- (2,0);
		\node at (3,0.5) {$\gamma k_1\gamma^{-1}$};
		\filldraw[black] (4,0) circle (2pt);
		\node at (4,-0.5) {$\sigma_{\gamma k_1\gamma^{-1}}$};
		\end{tikzpicture}
	\caption{}
	\label{boundkhomd}
	\end{subfigure}
	\begin{subfigure}{0.5\textwidth}
		\centering
		\begin{tikzpicture}
		\draw[very thick,->] (2,0) -- (1,0);
		\draw[very thick] (1,0) -- (0,0);
		\node at (1,0.5) {$\gamma k_1 k_2\gamma^{-1}$};
		\filldraw[black] (2,0) circle (2pt);
		\node at (2,-0.5) {$\sigma_{\gamma k_1k_2\gamma^{-1}}$};
		\end{tikzpicture}
	\caption{}
	\label{boundkhome}
	\end{subfigure}
	\caption{$K$ argument consistency.}
	\label{fig:boundkhom}
\end{figure}
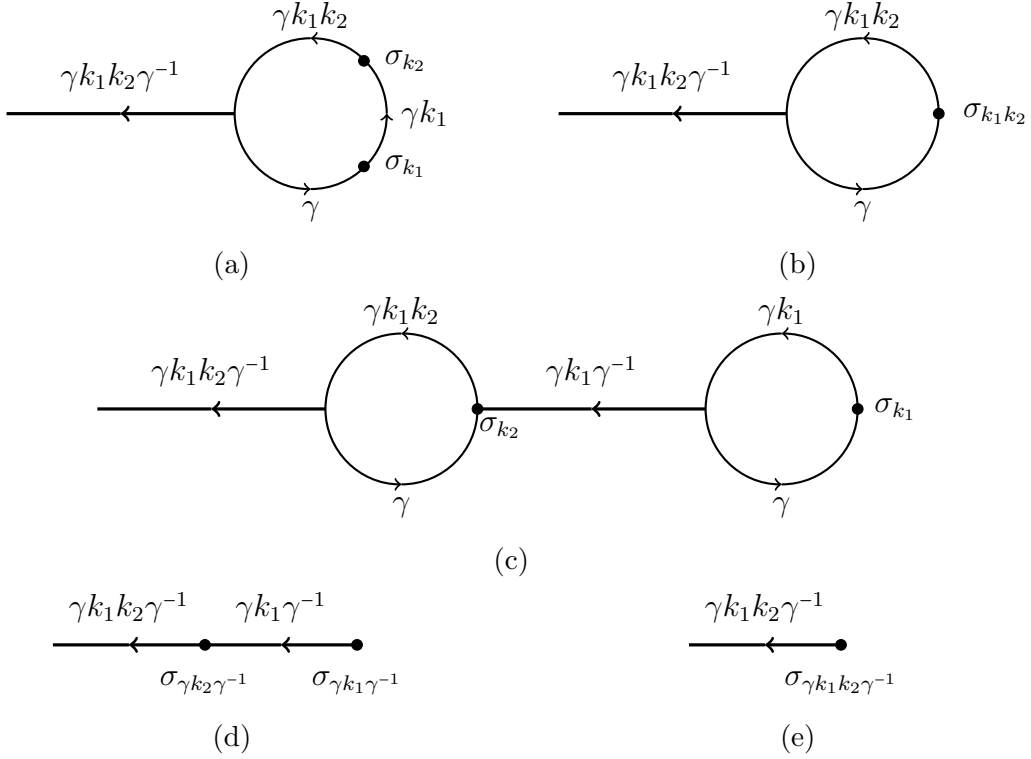

In order to constrain $B$ in its $K$ argument, we begin with the configuration shown in Figure~\ref{boundkhoma}.  We can then fuse the $\sigma_{k_1}$ and $\sigma_{k_2}$ TPOs to arrive at Figure~\ref{boundkhomb}.  Shrinking the circle then brings us to Figure~\ref{boundkhome}, while introducing a phase $B(\gamma,\gamma k_1k_2\gamma^{-1})$.  Alternatively, we could have deformed Figure~\ref{boundkhoma} to Figure~\ref{boundkhomc} by pinching off a circle containing $\sigma_{k_1}$.  Shrinking both circles in Figure~\ref{boundkhomc} leads to Figure~\ref{boundkhomd}, while producing the phase $B(\gamma,\gamma k_1\gamma^{-1})B(\gamma,\gamma k_2\gamma^{-1})$.  Finally, fusing the TPOs in Figure~\ref{boundkhomd} leads to Figure~\ref{boundkhome}.  Consistency then requires
\be
B(\gamma,\gamma k_1k_2\gamma^{-1})=B(\gamma,\gamma k_1\gamma^{-1})B(\gamma,\gamma k_2\gamma^{-1}),
\ee
matching (\ref{khomcond}).

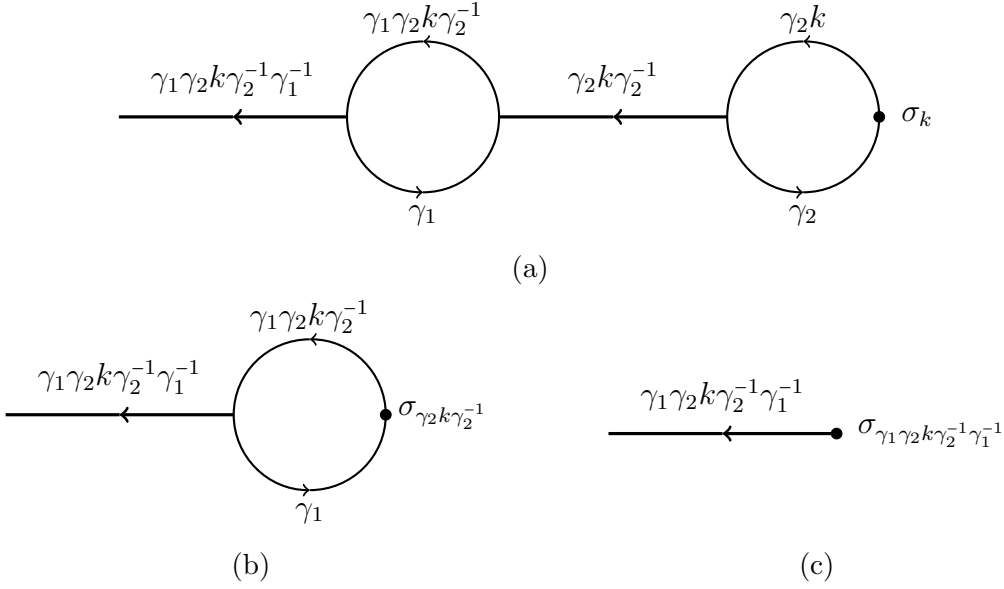
\begin{figure}
	\begin{subfigure}{\textwidth}
		\centering
		\begin{tikzpicture}
		\draw[very thick,->] (3,0) -- (1.5,0);
		\draw[very thick] (1.5,0) -- (0,0);
		\node at (1.5,0.5) {$\gamma_1\gamma_2 k\gamma_2^{-1}\gamma_1^{-1}$};
		\draw[thick,->] (4,0) [partial ellipse=-90:90:1cm and 1cm];
		\node at (4,-1.3) {$\gamma_1$};
		\draw[thick,->] (4,0) [partial ellipse=90:270:1cm and 1cm];
		\node at (4,1.3) {$\gamma_1\gamma_2k\gamma_2^{-1}$};
		\draw[very thick,->] (8,0) -- (6.5,0);
		\draw[very thick] (6.5,0) -- (5,0);
		\node at (6.5,0.5) {$\gamma_2 k\gamma_2^{-1}$};
		\draw[thick,->] (9,0) [partial ellipse=-90:90:1cm and 1cm];
		\node at (9,1.3) {$\gamma_2 k$};
		\draw[thick,->] (9,0) [partial ellipse=90:270:1cm and 1cm];
		\node at (9,-1.3) {$\gamma_2$};
		\filldraw[black] (10,0) circle (2pt);
		\node at (10.5,0) {$\sigma_k$};
		\end{tikzpicture}
	\caption{}
	\label{boundxhoma}
	\end{subfigure}
	\begin{subfigure}{0.5\textwidth}
		\centering
		\begin{tikzpicture}
		\draw[very thick,->] (3,0) -- (1.5,0);
		\draw[very thick] (1.5,0) -- (0,0);
		\node at (1.5,0.5) {$\gamma_1\gamma_2k\gamma_2^{-1}\gamma_1^{-1}$};
		\draw[thick,->] (4,0) [partial ellipse=-90:90:1cm and 1cm];
		\node at (4,-1.3) {$\gamma_1$};
		\draw[thick,->] (4,0) [partial ellipse=90:270:1cm and 1cm];
		\node at (4,1.3) {$\gamma_1\gamma_2 k\gamma_2^{-1}$};
		\filldraw[black] (5,0) circle (2pt);
		\node at (5.75,0) {$\sigma_{\gamma_2k\gamma_2^{-1}}$};
		\end{tikzpicture}
	\caption{}
	\label{boundxhomb}
	\end{subfigure}
	\begin{subfigure}{0.5\textwidth}
		\centering
		\begin{tikzpicture}
		\draw[very thick,->] (3,0) -- (1.5,0);
		\draw[very thick] (1.5,0) -- (0,0);
		\node at (1.5,0.5) {$\gamma_1\gamma_2 k\gamma_2^{-1}\gamma_1^{-1}$};
		\filldraw[black] (3,0) circle (2pt);
		\node at (4.25,0) {$\sigma_{\gamma_1\gamma_2 k\gamma_2^{-1}\gamma_1^{-1}}$};
		\end{tikzpicture}
		\vspace{1cm}
	\caption{}
	\label{boundxhomc}
	\end{subfigure}
	\caption{$\Gamma$ argument consistency.}
	\label{fig:boundxhom}
\end{figure}

For the $\Gamma$ argument, we begin with Figure~\ref{boundxhoma}.  Shrinking the rightmost circle would produce the TPO $\sigma_{\gamma_2 k\gamma_2^{-1}}$ and the phase $B(\gamma_2,\gamma_2 k\gamma_2^{-1})$, leading to Figure~\ref{boundxhomb}.  We can then shrink the remaining circle to produce Figure~\ref{boundxhomc} and pick up $B(\gamma_1,\gamma_1\gamma_2 k\gamma_2^{-1}\gamma_1^{-1})$.  Alternatively, we could have directly shrunk everything to the right of the $\gamma_1\gamma_2 k\gamma_2^{-1}\gamma_1^{-1}$ TDL in Figure~\ref{boundxhoma} to arrive directly at Figure~\ref{boundxhomc}, giving the phase $B(\gamma_1\gamma_2,\gamma_1\gamma_2k\gamma_2^{-1}\gamma_1^{-1})$.  Demanding equality of these paths leads to
\be
B(\gamma_1\gamma_2,\gamma_1\gamma_2k\gamma_2^{-1}\gamma_1^{-1})=B(\gamma_1,\gamma_1\gamma_2k\gamma_2^{-1}\gamma_1^{-1})B(\gamma_2,\gamma_2k\gamma_2^{-1}),
\ee
reproducing (\ref{xhomcond}) from the main text.

\section{Trivial Symmetries as a Mixed Extension}
\label{app:mixedext}

In \cite{Tachikawa:2017gyf}, Y.~Tachikawa explores the idea that higher-form symmetries of varying degrees can mix in generalizations of group extensions, i.e.~a theory could have a theory with symmetry $\Omega$ fitting into a short exact sequence
\be
\label{mnext}
1 \: \longrightarrow \: M_{[p]}
\: \longrightarrow \: \Omega
\: \longrightarrow \: N_{[q]} \longrightarrow \: 1,
\ee
where a subscript $[n]$ denotes an $n$-form symmetry.  The conjecture of \cite{Tachikawa:2017gyf} is that such an extension should be associated to a fibration of Eilenberg-Mac Lane spaces
\be
K(M,p+1)\longrightarrow B\Omega \longrightarrow K(N,q+1),
\ee
and that $B\Omega$ should be the classifying space for the mixed symmetry $\Omega$.

As this paper focuses on a mix of zero- and one-form symmetries, we might wonder if our framework can be cast in this language.  Qualitatively, we have seen that a theory with trivially-acting symmetries includes zero-form symmetries which can stand on their own, i.e.~they are proper subsymmetries of the system which we can gauge.  It also includes one-form symmetries (in the form of a non-trivial set of TPOs) which can not be separately gauged, as they are bound to the zero-form symmetries.  Examining (\ref{mnext}), this would be consistent with $p=0$ and $q=1$, which is to say our one-form symmetry is extended non-trivially by our zero-form symmetry.  
Matching the notation used elsewhere in the paper, 
then, we would formally regard a theory with zero-form symmetry $\Gamma$ and trivially-acting abelian subgroup $K$ as having a mixed symmetry described by $\Omega$ in\footnote{$K_{[1]}$ is also often written as $BK$, particularly in the mathematical literature.}
\be
\label{01ext}
1\longrightarrow \Gamma_{[0]} \longrightarrow \Omega  \longrightarrow K_{[1]} 
\longrightarrow 1.
\ee

Note that such an extension is opposite to the case which is most often considered in the literature, in which a higher-form symmetry extends a zero-form symmetry \cite{Yu:2020twi,Pantev:2022kpl,Benini:2018reh,Baez:2005sn,Wan_2019,Kobayashi_2019,KapustinThorngren}.

\subsection{Interpretation of the Extension Class}

In the following we focus on the case where $\Gamma$ is abelian, as that allows for the most concrete statements.  
Extensions of the form~(\ref{01ext}), interpreted in the category of complexes of abelian groups, induce\footnote{
We would like to thank T.~Pantev for results on such extensions.  Note in passing that the assignment of such homomorphisms to extensions~(\ref{01ext}) is natural, as one might suspect that they should correspond to elements of
\begin{equation}  
\label{extclass}
H^2_{\rm sing}( B^2 K, \Gamma) \: = \: {\rm Hom}( K(K,2), K(\Gamma,2) )
\: = \: {\rm Hom}(K,\Gamma),
\end{equation}
a computation described in e.g.~\cite[section 3.2.1]{Lin:2022xod}, \cite{stackexchange}.} a homomorphism
$K \rightarrow \Gamma$ (though not every such homomorphism may necessarily arise from an extension~(\ref{01ext})).

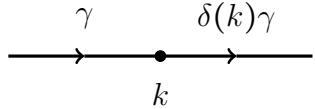
\begin{figure}[h]
	\centering
	\begin{tikzpicture}
	\draw[very thick,->] (0,0) -- (1,0);
	\draw[very thick] (1,0) -- (2,0);
	\draw[very thick,->] (2,0) -- (3,0);
	\draw[very thick] (3,0) -- (4,0);
	\node at (1,0.5) {$\gamma$};
	\node at (3,0.5) {$\delta(k)\gamma$};
	\filldraw[black] (2,0) circle (2pt);
	\node at (2,-0.5) {$k$};
	\end{tikzpicture}
	\caption{The general case of a point operator acting by a hom $\delta:K\to\Gamma$.}
	\label{fig:pointhom}
\end{figure}

As with group extensions, a non-trivial extension class should in general obstruct $K_{[1]}$ from being a subsymmetry of the system.  We can see this physically as follows.  Letting the extension class be $\delta: K\to\Gamma$, a point labeled by $k$ affects an incoming line labeled by some $\gamma$ as in Figure~\ref{fig:pointhom}.

\begin{figure}[h]
	\centering
	\begin{tikzpicture}
	\draw[very thick,->] (0,0) -- (1,0);
	\draw[very thick] (1,0) -- (2,0);
	\draw[very thick,->] (2,0) -- (3,0);
	\draw[very thick] (3,0) -- (4,0);
	\node at (1,0.5) {$\gamma$};
	\node at (3,0.5) {$\gamma$};
	\filldraw[black] (2,0) circle (2pt);
	\node at (2,-0.5) {$1$};
	\draw[dashed] (2,0) -- (2,1);
	\filldraw[black] (2,1) circle (2pt);
	\node at (2,1.5) {$k$};
	
	\draw[thick,->] (5.25,0.5) -- (5.75,0.5);
	
	\draw[very thick,->] (7,0) -- (9,0);
	\draw[very thick] (9,0) -- (11,0);
	\node at (9,-0.5) {$\gamma$};
	\filldraw[black] (9,1) circle (2pt);
	\node at (9,1.5) {$k$};
	\end{tikzpicture}
	\caption{When $\delta(k)=1$, the point operator labeled by $k$ can be `dragged off' of its line.  The resulting identity point and line operators can then be erased, leaving the point operator free floating.}
	\label{fig:pulloff}
\end{figure}
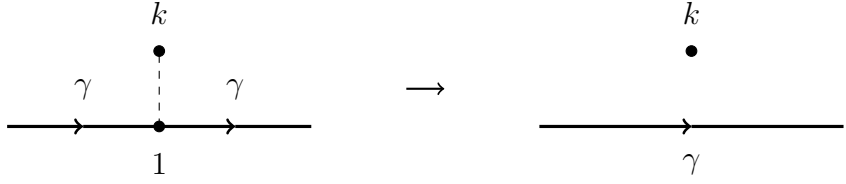

When $\delta$ is non-trivial, the $K$ points are ``bound'' to the $\Gamma$ lines.  If it happens that ker$(\delta)$ is all of $K$, however, we could alter Figure~\ref{fig:pointhom} as shown in Figure~\ref{fig:pulloff},
so a trivial extension class ``frees'' the points from the lines, allowing each symmetry to stand on its own.  In the cases examined in this paper, we always have $K$ as a normal subgroup of $\Gamma$, and so there is a natural candidate for $\delta$: the inclusion map.

\subsection{Orbifolds and Mixed Extensions}

Suppose we gauge $\Gamma_{[0]}$.  Given our assumption that the total symmetry of the system is an extension of $K_{[1]}$ by $\Gamma_{[0]}$, then from the formal considerations discussed so far, gauging $\Gamma_{[0]}$ should, on its face, bring us to a theory whose symmetry is an extension of $K_{[1]}$ by Rep$(\Gamma)_{[0]}$.

We will describe the theory both before and after gauging as completely as we can, still under the assumption that $\Gamma$ is abelian.  Our setup, as it has been before, is a 2d CFT with zero-form symmetry $\Gamma=K.G$, where $K$ acts trivially.  The total symmetry should be an extension of $K_{[1]}$ by $\Gamma_{[0]}$, with the extension class given by the inclusion map $\delta:K\to\Gamma$.  This map is injective, so $\delta$ has trivial kernel, and therefore all of the non-trivial TPOs are bound to lines.  The theory therefore has a unique vacuum and does not exhibit decomposition.

When we orbifold by $\Gamma_{[0]}$, if there is no mixed anomaly between the points and the lines, each TPO from the ungauged theory becomes a freestanding operator in the gauged theory.  The extension class $\tilde{\delta}$ of the gauged theory then has all of $K$ as its kernel.  The zero-form symmetry of the gauged theory is given by $\hat{\Gamma}$, the Pontryagin dual to $\Gamma$.  In order to figure out what mixed anomaly the gauged theory has, we can note that $\hat{\Gamma}$ acts on local operators through phases given by characters:
\be
\hat{\gamma}\cdot\sigma_k
\: = \:
\sigma_k\times\chi_{\hat{\gamma}}(\delta(k)),
\ee
from which we see that the mixed anomaly is given by $\chi_{\hat{\gamma}}(\delta(k))$.

In fact, we can repeat this argument to learn more about the mixed anomaly in the ungauged theory.  Note that gauging $\hat{\Gamma}$ brings us to a theory with symmetry $\hat{\hat{\Gamma}}$, which by Pontryagin duality is canonically isomorphic to $\Gamma$.  This is the sense in which gauging the quantum symmetry $\hat{\Gamma}$ undoes the first gauging to return us to the original theory.  But we can repeat the argument from earlier -- the mixed anomaly in the doubly gauged theory should be given by a character of $\hat{\hat{\Gamma}}$ as $\chi_{\hat{\hat{\gamma}}}(\tilde{\delta}(k))$.  Of course we can map this by isomorphism to a character of $\Gamma$, so we expect that the mixed anomaly in the ungauged theory can be written as $\chi_\gamma(\tilde{\delta}(k))$.  For ease of comparison, let us summarize these conclusions in tabular form:

\begin{center}
	\begin{tabular}{| l | l | l |}
		\hline
		& Ungauged Theory & Gauged Theory\\
		\hline
		Total Symmetry & $1\to\Gamma_{[0]}\to\Omega\to K_{[1]}\to 1$ & $1\to\hat{\Gamma}_{[0]}\to\tilde{\Omega}\to K_{[1]}\to 1$ \\
		Extension Class & $\delta$ &  $\tilde{\delta}$ \\
		Mixed Anomaly & $\chi_\gamma(\tilde{\delta}(k))$ & $\chi_{\hat{\gamma}}(\delta(k))$ \\
		\hline
	\end{tabular}
\end{center}

Note how the extension class in the ungauged theory determines the mixed anomaly in the gauged theory and vice versa.  This is entirely in line with the analysis of \cite{Tachikawa:2017gyf}, which predicts that this phenomenon should arise any time we gauge subsymmetries fitting into extensions.

These results are also in line with the predictions of decomposition.  For instance, it is well known that if we gauge a trivially-acting abelian group $K$, we end up with a direct sum of $|K|$ copies of the original theory.  From the topological operator point of view, the vacuum in each theory is a local topological operator which originated as one of the TPOs that implemented the trivial symmetry in the ungauged theory (or possibly a linear combination of those TPOs).  When we supplement a theory with a trivially-acting symmetry that has non-trivial mixed anomaly, the table above suggests that we should find a non-trivial extension class in the gauged theory.  This means that not all of the TPOs in the gauged theory will be freestanding -- decomposition will be obstructed (relative to the same theory without mixed anomaly).  It also means that the resulting theory should have its own trivially-acting symmetries implemented by the TPOs that are still bound to lines.  The decomposition structure of orbifolds by non-effective symmetries with non-trivial mixed anomaly was investigated (under the name of quantum symmetry phases) in \cite{Robbins:2021lry,Robbins:2021ibx,Robbins:2021xce}, and indeed turning on a mixed anomaly obstructs decomposition in the gauged theory (as can also be seen from the examples in section~\ref{sec:examples}).

If we were to attempt to extend this analysis to non-abelian $\Gamma$, we would quickly run into trouble.  The calculation (\ref{extclass}) of the extension class only makes sense when $\Gamma$ is abelian, so while it is conceivable that we still have a classifying space given as a fibration, that fact alone would not get us far.  In this case the approach taken in the main text where we focus on the physical behavior of the topological operators seems preferable, as that tactic allows us to work out the behavior of non-abelian (and even non-group-like) examples.

\newpage

\addcontentsline{toc}{section}{References}

\bibliographystyle{utphys}
\bibliography{TopOps}

\end{document}